\documentclass[aps,prl,reprint]{revtex4-2}
\usepackage{amsmath}
\usepackage{amssymb}
\usepackage{graphicx}
\usepackage{mathrsfs}
\usepackage{hyperref}
\usepackage{braket}
\usepackage[caption=false]{subfig}
\usepackage[dvipsnames]{xcolor}

\begin{document}

\title{Soliton resuscitations: asymmetric revivals of the breathing mode of an atomic bright soliton in a harmonic trap}

\author{Waranon Sroyngoen}
\author{James R.~Anglin}

\affiliation{\mbox{State Research Center OPTIMAS and Fachbereich Physik,} \mbox{Technische Univerit\"at Kaiserslautern,} \mbox{D-67663 Kaiserslautern, Germany}}

\date{\today}

\begin{abstract}
An atomic bright soliton realised in a quasi-one-dimensional Bose-Einstein condensate can be considered as an open quantum system. The soliton's breathing mode, for example, is damped by emission of atoms from the soliton to spatial infinity, which thus acts as a Markovian environment for the soliton. If the soliton is held in a shallow harmonic trap, however, the environment becomes non-Markovian: emitted atoms oscillate in the trap and eventually return to the soliton, interfering with it, producing periodic revivals of the breathing mode (``resuscitations''). 
The amplitude envelopes of these breathing revivals shows a curious asymmetry, with a gradual increase in breathing amplitude followed by sudden drop in amplitude that becomes more and more pronounced in later revivals. We explain this asymmetrical revival pattern in the non-Markovian revivals by deriving a close analytical approximation to the Bogoliubov-de Gennes frequency spectrum for the weakly trapped soliton.
\end{abstract}

\maketitle

\section{I. Introduction}
\subsection{I.A. Bright solitons} Bright solitons are named for their appearance as self-focused light pulses in nonlinear optical fibers\cite{Fibers}, but they can also occur in condensed Bose gases\cite{Condensates,Trains}, as self-binding concentrations of atomic vapor rather than of light. Similar solitons occur in both kinds of medium, because bright solitons are solutions to the non-linear Schr\"odinger equation in one spatial dimension with a negative nonlinear term. When the nonlinear Schr\"odinger equation appears as the Gross-Pitaevskii equation for the macroscopic wave function of a dilute Bose-Einstein condensate, a negative scattering length for interatomic collisions provides the necessary negative nonlinearity for bright solitons. In order to support such solitons, the macroscopic wave function $\psi(x,t)$ of the condensate must effectively be confined to only one spatial dimension. 

Experiments have produced such atomic bright solitons in Lithium\cite{Condensates,Trains,Evaporation}, Potassium\cite{Potassium}, and Cesium\cite{Cesium} vapors. Further experiments have observed reflections\cite{Reflect} and collisions\cite{Collide} of atomic bright solitons. An atomic Mach-Zehnder interferometer using bright solitons has been shown to have enhanced fringe visibility\cite{Interferometer}. 
Trains of multiple bright solitons\cite{Trains}, and collective motional modes of the trains\cite{BreathingTrain}, have also been studied.

\subsection{I.B. Collective ``breathing'' excitations}
Collective excitations of individual solitons have attracted interest as well, and indeed the distinction between collective motional modes of multi-soliton groups, and the vibration-like excitation modes of individual solitons, is not always sharp, because in some cases the multiple solitons never fully separate from each other. Higher-order ``breather'' solitons\cite{BreatherRelaxation,BreatherBreakup,BreatherFluctuation,Breathers} can be considered single solitons with nonlinearly large-amplitude breathing excitations, or they can be seen as interactions of two or more solitons that repeatedly merge and (perhaps only partially) separate. Quantum effects in these nonlinear excitations, for the dilute condensed Bose gas beyond the mean-field nonlinear Schr\"odinger equation, have been studied theoretically using truncated Wigner\cite{BreatherRelaxation}, Bethe Ansatz\cite{BreatherBreakup} and Bogoliubov-de Gennes theory\cite{BreatherFluctuation}.

The low-amplitude, linear limit of a single-soliton breathing mode, as illustrated in Fig.~1, has also been induced in experiments\cite{Breathing}, with the breathing clearly observed over many periods.\begin{figure}[!h]
    \centering
\includegraphics[width=0.45\textwidth]{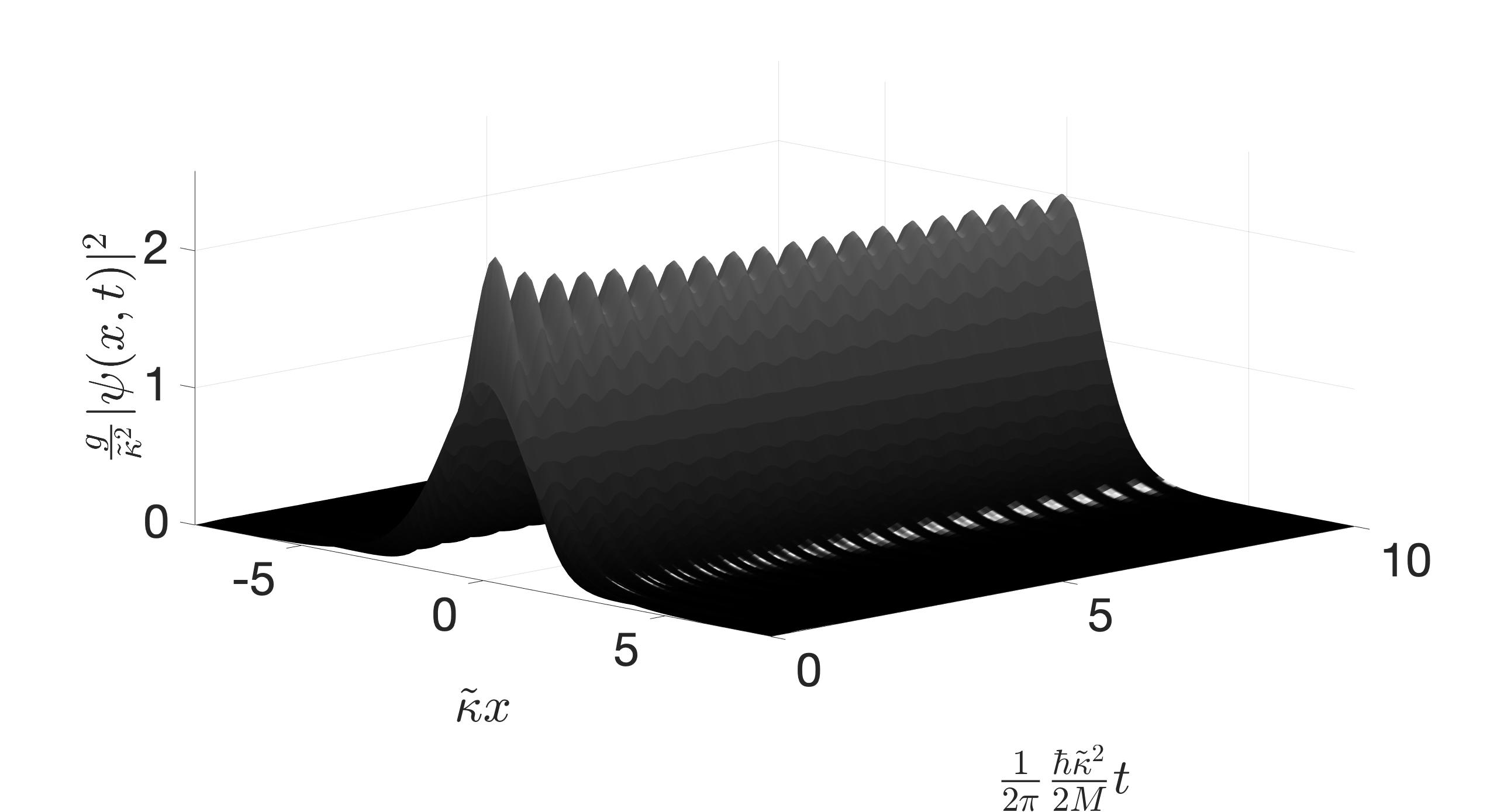}
    \caption{Small-amplitude soliton breathing in mean field theory, with no trapping potential. The gas density in the attractively interacting condensate is self-bound in a $\mathrm{sech}^2$ profile; the width and peak density of the spatial profile oscillate periodically in time, while the oscillation amplitude decays due to dispersion.}
\end{figure}

\begin{figure*}[!t]
    \includegraphics[width=\textwidth]{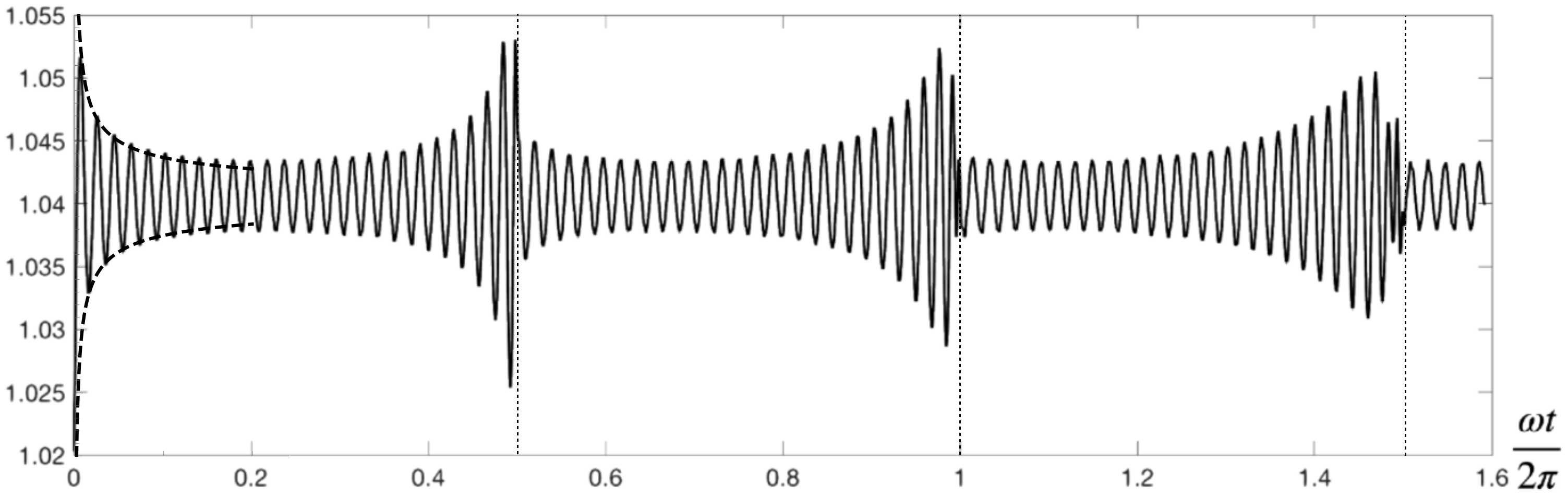} 
    \caption{\label{trump}Asymmetrical resuscitations of soliton breathing in a harmonic trap of frequency $\omega$, as shown in the peak condensate density $|\psi(0,t)|^2$ as a function of time in trap periods. The evolution is given by the one-dimensional Gross-Pitaevskii equation from the non-stationary initial wave function $\psi(x,0)=A(\tilde{\kappa}/\sqrt{g})\mathrm{sech}(\tilde{\kappa} x)$  for $\tilde{\kappa} = \sqrt{50}a_0^{-1}$, where $a_0=\sqrt{\hbar/(2M\omega)}$ (with atomic mass $M$) is the harmonic trap width. The case $A=1$ would be a perfect bright soliton, with no breathing; in this case $A=1.01$, providing a small-amplitude breathing excitation at the chemical potential frequency $\hbar\kappa^2/(2M)=A^4(\tilde{\kappa} a_0)^2\omega\doteq 52\omega$. The amplitude initially decays in an envelope that closely fits the $t^{-1/4}$ steepest descents approximation of Eqn.~(\ref{B0SD}) (dashed curves), but then the breathing amplitude revives: it recovers to a maximum amplitude approximately every half-period of the harmonic trap. In fact the first revival begins noticeably \emph{sooner} than a trap half-period; note as well the ``trumpet-shaped'' asymmetrical envelopes of the revived oscillations.}
\end{figure*}
This linear breathing is not a unique normal mode, however. Small-amplitude excitations of a single bright soliton with no external trapping potential have a continuous frequency spectrum\cite{Kaup}, and half of all modes (the ones with even spatial parity) include some component of ``breathing'' (oscillation in the soliton  width and amplitude). Although there is thus no unique breathing mode, and no unique breathing mode frequency, typical initial state preparation mainly excites low-frequency perturbations of the background soliton, and interference between the background order parameter and low-frequency perturbations makes the gas density breathe at the chemical potential frequency, while dispersion in the excitation continuum makes the breathing amplitude gradually decay, within a $t^{-1/2}$ envelope\cite{Root-t}, as is visible in Fig.~1.

For a breathing soliton which is purely self-trapped, with no external potential, part of this dispersive relaxation process includes emission of small-amplitude waves to infinity. For atomic bright solitons described in Gross-Pitaevskii mean field theory, this implies ejection of atoms from the soliton, which carry away energy as well as particles from it forever. Infinity---or simply spatial regions well outside the soliton's width---thereby act as an environment within which the soliton moves or breathes as an open system.

\subsection{I.C. Breathing solitons in traps}
Bright solitons in experiments, however, are generally held in trapping potentials that can be approximated well as harmonic. Ejected atoms will therefore not escape to infinity, but rather oscillate in the trap and return to the soliton. If they retain phase coherence with the macroscopic wave function of the soliton, these returning atoms will interfere with the soliton, and the breathing mode will therefore revive periodically. Fig.~2 shows an example of such breathing revivals---``resuscitations''---as predicted by mean field theory (\emph{i.e.} under the time-dependent Gross-Pitavskii nonlinear Schr\"odinger equation in one spatial dimension).
 
These resuscitations of the soliton breathing mode are thus a simple example of non-Markovian open-system evolution, in which the environment effectively remembers information that has been received from the system, so that its present effects on the system depend on the system's prior history. In this case the non-Markovian nature of the environment is simply oscillation of atoms in a harmonic potential, and so the most striking feature of the resuscitations is simply that they recur at every half-period of the harmonic trap. Looking more closely at Fig.~2, however, we notice the curiously asymmetric, trumpet-shaped envelopes of these resuscitations, which seem to become more and more asymmetric in each subsequent revival. Even such a simply non-Markovian environment as a harmonic trap can produce complex open-system evolution, through interplay between the non-Markovian environment and non-trivial system dynamics, such as the breathing of a bright soliton with attractive mean-field interactions.

In this paper we analytically study these temporally asymmetrical soliton breathing revivals, by solving the linearized mean-field evolution of a soliton in a shallow harmonic trap using matched asymptotics in the Bogoliubov-de Gennes equations. Our results illustrate the complexity of non-Markovian evolution even in a simple instance, and they will also provide a mean-field background against which quantum many-body effects may be discerned in future studies and experiments.

\subsection{I.D. Paper structure}
Our paper is structured to avoid burdening readers with technical details before the meaning of the details is clear. In Section II we briefly review the untrapped atomic bright soliton in Gross-Pitaevskii mean field theory, as well as its linearized Bogoliubov-de Gennes perturbations. In Section III we then briefly present our results for the modified Bogoliubov-de Gennes modes in the harmonic trap, without derivation, and in Section IV we show how these results lead to the asymmetrical resuscitations that we have described, before concluding and discussing the outlook for future study in Section V. Technical details, including the derivation by matched asymptotics of the main results that we applied in Section III, are then provided in four rather lengthy appendices.

\section{II. The Untrapped Bright Soliton}
\subsection{II.A. The mean field background}
We consider a quasi-one-dimensional Bose-Einstein condensate with attractive interactions, using Gross-Pitaevskii mean field theory with Bogoliubov-de Gennes linearized excitations. As we note in Appendix A, mean field theory can be expected to be highly accurate for slightly excited atomic bright solitons, even with modest condensate numbers, because the \emph{quantum depletion}
\begin{equation}
    \langle\hat{\psi}^\dagger(x)\hat{\psi}(x)\rangle - |\Psi_0|^2
\end{equation}
of a stationary atomic bright soliton is only one-third: not one-third of the condensate, but \emph{one-third of an atom}. As the zeroth-order background state of the condensate's macroscopic wave function $\psi(x,t)$, we therefore assume the bright soliton form
\begin{equation}\label{alpha_1}
    \Psi_0(x,t) = e^{i\frac{\hbar\kappa^2}{2M}t}\frac{\kappa}{\sqrt{g}} \mathrm{sech}(\kappa x)\;,
\end{equation}
where $-g$ is the one-dimensional scattering strength for collisions between the condensate particles, $M$ is the particle mass, and $\kappa$ is a constant inverse length. In one dimension, the scattering strength $g$ is an inverse length; for quasi-one-dimensionally trapped gases, $g$ is proportional to the three-dimensional scattering length divided by the mean cross section of transverse confinement. Assuming here negative scattering length, we consider only $g>0$.

The background $\Psi_0$ of (\ref{alpha_1}) solves the Gross-Pitaevskii nonlinear Schr\"odinger equation, which with our definition of $g$ reads as
\begin{equation}\label{alpha_2}
    i\hbar\frac{\partial}{\partial t}\psi = -\frac{\hbar^2}{2M}\frac{\partial^2}{\partial x^2}\psi + V(x)\psi- \frac{\hbar^2 g}{M}|\psi|^2\psi\;
\end{equation}
in general; the solution $\psi\to\Psi_0$ as given by (\ref{alpha_1}) applies exactly in the special case $V(x)=0$, but also offers a good approximate solution for any $V(x)$ that varies slowly over the soliton width scale $\kappa^{-1}$.

The inverse length $\kappa$ is arbitrary, inasmuch as it depends on how many atoms are in the Bose-condensed gas. Conversely, for given $\kappa$ the condensate atom number in the bright soliton is
\begin{equation}\label{number}
N_0 = \int\!dx\,|\Psi_0|^2 = \frac{\kappa}{g}\int\!d(\kappa x)\,\mathrm{sech}^2(\kappa x) = \frac{2\kappa}{g}\;.
\end{equation}

\subsection{II.B. Bogoliubov-de Gennes perturbations}

For small, time-dependent perturbations around the bright soliton background, \begin{equation}\label{Psiex}\psi(x,t) = \Psi_0(x,t)+e^{i\frac{\hbar\kappa^2}{2M}t}\delta\psi(x,t)\;,\end{equation} the perturbation $\delta\psi$ can be expanded in a set of normal modes:
\begin{align}\label{BdG}
    \delta{\psi}(x,t) =& \int_{-\infty}^\infty\!dk\,\Big[{\beta}_k e^{-i\Omega(k)t}u_k(x) + {\beta}^*_k e^{i\Omega(k)t}v^*_k(x)\Big]\nonumber\\
    \Omega(k) =& \frac{\hbar}{2M}(k^2+\kappa^2)
\end{align}
for arbitrary (small) complex constants $\beta_k$. In addition to this continuous set of $(u_k,v_k)$ excitations, the bright soliton also possesses two discrete zero modes; as we discuss in Appendix B, these modes are not directly involved in the breathing excitations that we discuss in this paper. One of them is indirectly involved, in that the requirement to ``fix'' this zero mode will resolve an ambiguity in distinguishing which part of our initial $\psi(x,0)$ is the soliton background and which part is the perturbation.

To make this perturbed $\psi(x,t)$ solve the Gross-Pitaevskii equation (\ref{alpha_2}) to first order in $\delta\psi$, particular functional forms are required for the mode functions $u_k(x)$ and $v_k(x)$. These forms are known explicitly and exactly for the case $V=0$ \cite{Kaup}, and it will be a major goal in this paper to find them approximately for a harmonic $V$. For $V=0$, we have the normalized mode functions that are\cite{Kaup}\begin{widetext}
\begin{align}\label{Kaup}
    u_k(x) &= \frac{1}{\sqrt{2\pi}}\frac{\kappa^2}{k^2+\kappa^2}\left(\frac{k^2-\kappa^2}{\kappa^2}+\frac{2ik}{\kappa}\tanh(\kappa x)+\mathrm{sech}^2(\kappa x)\right)e^{ikx}\nonumber\\
    v_k(x) &=\frac{1}{\sqrt{2\pi}}\frac{\kappa^2}{k^2+\kappa^2}\mathrm{sech}^2(\kappa x)e^{ikx}\;.
\end{align}\end{widetext}

The spectrum of $k$ is indeed continuous, and no $u_k$ actually decay at large $|x|$: the bright soliton does not have a discrete spectrum of collective multipole excitation modes, as a condensate held in an external trap does. Long-wavelength perturbations, with amplitude concentrated in $|k|\ll\kappa$, will translate the soliton, if they have odd spatial parity; if they have even parity, they can make the soliton ``breathe''. Otherwise, if there are initial perturbations present that change the condensate profile away from the perfect soliton's $\mathrm{sech}$ profile in more complicated ways than translation or breathing, then these excitations will have to have significant amplitude at $|k|\gtrsim \kappa$, in order to affect $\psi(x,0)$ on wavelengths shorter than the soliton width. All such initial excitations will immediately disperse away from the soliton at speeds of order $\hbar\kappa/M$: they represent pure transients, and not longer-lived collective excitations of the soliton. So, apart from translation, breathing is actually the only thing like a collective mode which a bright soliton can possess. 

In this paper, therefore, we will analyse only the particular kind of small perturbations around $\Psi_0$ that evolve from slightly stretched or compressed initial conditions of the form
\begin{equation}\label{Ini}
    \psi(x,0) = (1+\delta)\frac{\tilde{\kappa}}{\sqrt{g}} \mathrm{sech}(\tilde{\kappa} x)=\Psi_0(x,0)+\delta\psi(x,0)\;,
\end{equation}
where $\delta$ is small, and where there is a subtle distinction between the $\tilde{\kappa}$ scale that appears in our initial condition, and the corresponding scale $\kappa$ in our soliton background. 

If $\delta>0$ then the initial wave function is slightly too high, in comparison to its width, to be a perfect $\Psi_0$ soliton with $\kappa\to\tilde{\kappa}$; if $\delta<0$ then $\psi(0,0)$ is not quite high enough. Our $\psi(x,0)$ is thus certainly not a perfect bright soliton, and we will consider $\psi(x,0) = \Psi_0(x,0)+\delta\psi(x,0)$. This leaves an ambiguity, however, in identifying which value of $\kappa$ we should use in $\Psi_0(x,0)$, since choosing a slightly different value of $\kappa$ would make only a small change in $\Psi_0$, and this small difference could simply be considered part of $\delta\psi(x,0)$.

As we show in Appendix B, however, requiring that $\delta\psi(x,t)$ should remain small over long times, and not grow slowly but steadily larger with $t$, selects one particular value for $\kappa$ for a given $\tilde{\kappa}$ and $\delta$:
\begin{equation}\label{kappafix}
    \kappa = (1+\delta)^2\tilde{\kappa}\;.
\end{equation}
This provides the unambiguous initial perturbation
\begin{align}\label{deltabr}
    \delta\psi(x,0) &= \frac{1}{1+\delta}\frac{\kappa}{\sqrt{g}}
    \mathrm{sech}\Big(
    \frac{\kappa x}{(1+\delta)^2}\Big)
    - \frac{\kappa}{\sqrt{g}}\mathrm{sech}(\kappa x)\nonumber\\
    &= \delta\frac{\kappa}{\sqrt{g}}[2\kappa x\tanh(\kappa x)-1]\mathrm{sech}(\kappa x) + \mathcal{O}(\delta^2)\;.
\end{align}

\subsection{II.C Breathing evolution}
When our bright soliton is initially perturbed by a small $\delta\psi(x,0)$ of the form (\ref{deltabr}), then, we can use the orthonormality of the $u_k(x),v_k(x)$ of (\ref{Kaup})
\begin{align}
    \int_{-\infty}^\infty\!dx\, (u_k u_{k'}^* - v_k v_{k'}^*) &= \delta(k-k')\nonumber\\
    \int_{-\infty}^\infty\!dx\, (u_k v_{k'} - v_k u_{k'})&=0
\end{align}
to obtain
    \begin{align}\label{betak1}
    \beta_k &= \int_{-\infty}^\infty\!dx\,\Big(u_k^*(x)\delta\psi(x,0) - v_k^*(x)\delta\psi^*(x,0)\Big)\nonumber\\
    &=-\sqrt{\frac{\pi}{2g}}\,\delta\,\mathrm{sech}\Big(\frac{\pi k}{2\kappa}\Big)\;,
\end{align}
where the exact evaluation of the integral is obtained through integration by parts followed by contour integration. Note that $\beta_{-k}=\beta_k$, so that the reflection symmetry of our initial state is preserved through time evolution.

With this specific form (\ref{betak1}) of $\beta_k$, we can integrate (\ref{BdG}) to obtain the breathing perturbation $\delta\psi(x,t)$ for all $t$. For small $\delta$, the density expanded to first order in $\delta$
\begin{equation}\label{firstorder}
    |\psi(x,t)|^2 \doteq |\Psi_0|^2 + \Psi_0\delta\psi^* + \Psi_0^*\delta\psi 
\end{equation}
is an excellent approximation to the exact $|\psi(x,t)|^2$ over many breathing periods. A simple measure of the overall amplitude of the breathing mode is then $|\psi(0,t)|^2$, which from (\ref{BdG}), (\ref{Kaup}), and (\ref{betak1}) is given by
\begin{align}\label{breathe}
    \frac{g}{\kappa^2}|\psi(0,t)| &= 1- \delta\  B_0(t) + \mathcal{O}(\delta^2) \nonumber\\
    B_0(t)& = \int_{-\infty}^\infty\!\frac{dk}{\kappa}\,\mathrm{sech}\Big(\frac{\pi k}{2\kappa}\Big)\cos\Big(\frac{\hbar(k^2+\kappa^2)t}{2M}\Big) \;.
\end{align}
For later times, the integral that defines the dimensionless breathing amplitude $B_0(\tau)$ in (\ref{breathe}) can be evaluated using the method of steepest descents to reveal the asymptotic behavior
\begin{equation}\label{B0SD}
    B_0(t)=\sqrt{\pi}\frac{\cos\Big(\frac{\hbar\kappa^2}{2M}t+\frac{\pi}{4}\Big)}{\sqrt{\frac{\hbar\kappa^2}{2M}t}}+\mathcal{O}(t^{-3/2})\;. 
\end{equation}
The breathing at the chemical potential frequency $\hbar\kappa^2/(2M)$ which we have described thus decays in amplitude as $t^{-1/2}$, due to dispersion.

Figure~1 shows an example of this breathing in the absence of a trapping potential, for an initial state (\ref{hotfix1}) with $\delta = 0.2$. The breathing that occurs from this initial state, under Gross-Pitaevskii mean-field evolution, does indeed have the frequency
\begin{equation}
    \omega_B = \frac{\hbar\kappa^2}{2M} = (1+\delta)^4\frac{\hbar\tilde{\kappa}^2}{2M}\;.
\end{equation}
In this illustrative case $\delta = 0.2$ is small enough that the Bogoliubov-de Gennes linearization of the perturbation is still quite accurate, but the quartic $(1+\delta)^4\doteq 2.1$ factor in $\omega_B$ is large enough to confirm that the soliton itself really ``knows'' that $\kappa$, rather than $\tilde{\kappa}$, is its correct background soliton width: There are nearly twenty-one breathing periods in the time plotted, not just ten.

Because the Bogoliubov-de Gennes breathing oscillation dies away from dispersion, at very late times the post-Bogoliubov-de Gennes $\mathcal{O}(\delta^2)$ corrections to $\psi$ may be more significant locally than the leading-order term (\ref{firstorder}). We can actually identify the main post-Bogoliubov-de Gennes effect, however, using only the leading order $\delta\psi(x,t)$ from (\ref{BdG}) with the coefficients $\beta_k$ from (\ref{betak1}).

\subsection{II.D Particle emission}
The reason we can find $\mathcal{O}(\delta^2)$ effects using only our $\delta\psi$ to order $\delta$ is that for $|x|\gg 1/\kappa$, the soliton background $\Psi_0(x,t)$ is exponentially small. In the region of large $x$, therefore, the full density is $|\delta\psi(x,t)|^2+\mathcal{O}(\delta^3)$. We can thus compute the number of particles which have moved far away from the soliton at time $t$ as
\begin{equation}
    \Delta N(t) = \int_L^\infty\!dx\,\Big(|\delta\psi(x,t)|^2+|\delta\psi(-x,t)|^2\Big)
\end{equation}
for any $L\gg 1/\kappa$. 

Evaluating $\delta\psi(x,t)$ for $|x|\geq L$ and at late $t$ using steepest descents, this lets us obtain
\begin{equation}
    \lim_{t\to\infty}\Delta N(t) = \delta^2 N_0 + \mathcal{O}(\delta^3,t^{-1})\;.
\end{equation}
This lets us know, without actually calculating the $\mathcal{O}(\delta^2)$ correction to $\delta\psi$, that the number of particles remaining in the soliton itself must gradually decrease over time, as the breathing mode decays. Mathematically we can recognize that the breathing mode decays due to dispersion, but the excitation energy is physically carried away from the soliton by particles that are emitted from the soliton to infinity.

This shaking loose of particles as the soliton breathes suggests the main difference which we can expect in the breathing of a soliton in a weak harmonic trap, compared to a soliton with no trap. As long as the trap is sufficiently weak, the soliton's breathing oscillation will initially be the same as without the trap, and so will the initial stages of its $1/\sqrt{t}$ dispersive decay. In the trap, however, particles which are shaken loose from the oscillating soliton will not be able to escape to infinity: instead they will return to the soliton, after half a trap period, and interfere again with $\Psi_0$. The breathing mode will resuscitate.

\section{III. Collective modes of a bright soliton in a shallow trap}
\subsection{III.A: The shallow trap}

Our goal is to understand how the breathing mode evolves when the condensate is held in a harmonic confining potential $V(x) \to \frac{M\omega^2}{2}x^2$. To maintain the open-system picture in which the soliton as a self-interacting system is only weakly or slowly affected by a long-range environment, we will focus on the limit in which the harmonic trap, with trap width $a_0=\sqrt{\hbar/(2M\omega)}$, is very broad and shallow compared to the bright soliton's width $1/\kappa$:
\begin{equation}\label{epsdef}
\frac{1}{\kappa a_0}=\frac{1}{\kappa}\sqrt{\frac{2M\omega}{\hbar}} =: \varepsilon \ll 1\;.
\end{equation}
In this limit the $V\to0$ bright soliton $\Psi_0(x,t)$ is still a solution to the Gross-Pitaevskii equation with the harmonic trap $V\not=0$, up to corrections that are only of order $\varepsilon^4$, which we will consistently neglect throughout this paper. 

\subsection{III.B: Discrete spectrum}
In this small-$\varepsilon$ limit the $u_k(x)$ and $v_k(x)$ modes of (\ref{Kaup}) are also still solutions to the Bogoliubov-de Gennes solutions, locally for $|x|\ll a_0$. Out of the continuum of $k$ modes that appear in the untrapped problem, however, only a particular discrete set of $k_n$ (for whole-number $n$) can be extended into solutions for all $x$. This will imply the discrete set of eigenfrequencies
\begin{align}
\delta\psi(x,t) &= e^{i\frac{\hbar \kappa^2}{2M}t}\sum_n \left( \beta_n e^{-i\Omega_n t}u_n(x) + \beta^*_n e^{i\Omega_n t}v_n^*(x)\right)\nonumber\\
    \Omega_n &= \frac{\hbar k_n^2}{2M}\;. 
\end{align}

The discrete set of modes will have spatial parity given by $n$,
\begin{equation}
    \left(\begin{matrix}u_n(x)\\ v_n(x)\end{matrix}\right) \underset{|x|\ll a_0}{\longrightarrow} Z_n \left[\left(\begin{matrix}u_{k_n}(x)\\ v_{k_n}(x)\end{matrix}\right)+(-1)^n\left(\begin{matrix}u_{-k_n}(x)\\ v_{-k_n}(x)\end{matrix}\right)\right]
\end{equation}
and will have finite normalization
\begin{equation}
\int\!dx\,(u^*_m u_n - v^*_m v_n) = \delta_{mn}
\end{equation}
ensured by the normalization factors $Z_n$.

\subsection{III.C: Results from matched asymptotics}
\subsubsection{Discrete frequencies}
In Appendix C we exploit the smallness of $1/\kappa a_0 = \varepsilon$ to apply the method of matched asymptotics, and construct the complete set of discrete $(u_n,v_n)$ by smoothly matching $(u_k,v_k)$ solutions from (\ref{Kaup}) for $|x|\lesssim \varepsilon^{1/2}a_0$ onto parabolic cylinder functions for $|x|\gtrsim \varepsilon^{1/2}a_0 = \varepsilon^{-1/2}\kappa^{-1}$. This lengthy analysis in Appendix C supplies the key results of our paper:
\begin{align}
k_n &= \varepsilon\kappa\sqrt{n+\frac{1}{2}+\Delta_n}\label{kn}\\
\Omega_n &= \frac{\hbar\kappa^2}{2M}+\omega\Big(n+\frac{1}{2}+\Delta_n\Big)\label{omegan}\\
\Delta_n &= \frac{2}{\pi}\cos^{-1}\left(\frac{1-\varepsilon^2n}{1+\varepsilon^2n}\right)\label{deltan3}\;.
\end{align}

Equation (\ref{deltan3}) says that $\Delta_n$ rises smoothly from 0 for small $n$ through 1 around $n\simeq \varepsilon^{-2}$, and ultimately approaches 2 for $n\to\infty$. The fact that $\Delta_{n\to\infty}=2$ means that $\nu_{N-2}=N$ for $N\to\infty$, so that the same large frequency range which holds $N\gg 1$ harmonic oscillator eigenfrequencies only holds $N-2$ of the bright soliton Bogoliubov-de Gennes eigenfrequencies.

The missing two eigenfrequencies are the zero mode and the Kohn dipole mode, which are discussed in Appendix B and whose frequencies lie below the $\hbar\kappa^2/(2M)$ gap. We could consider them to be the cases $n=-2$ and $n=-1$, respectively, since the zero mode is spatially even and the dipole mode is odd. Since the bright soliton amplitude $\kappa$ could in principle be raised continuously from zero, without ever altering the extremely high-frequency end of the excitation spectrum except by the overall chemical potential shift, the total number of eigenfrequencies below some high upper bound $\Omega_N = \hbar\kappa^2/(2M)+\omega(N+1/2)$, including the two modes below the $\hbar\kappa^2/(2M)$ gap, must be $N$ for all solitons. Pushing the zero mode and Kohn mode down below the gap means that the $N-2$ modes that remain above the gap are smoothly stretched out in frequency to cover the same total frequency range with two fewer eigenfrequencies. 
 
This subtle distortion of the Bogolibov-de Gennes eigenspectrum, in comparison to the evenly spaced harmonic oscillator spectrum, will be responsible for the main effect that we discuss in this paper. The fact that all the even-mode eigenfrequencies $\Omega_{2n}$ are spaced nearly $2\omega$ apart is responsible for the revivals of soliton breathing after every half-period $\pi/\omega$. The fact that all the intervals between successive $\Omega_{2n}$ are actually not quite exactly $2\omega$, but instead are slightly and systematically stretched according to (\ref{deltan3}) and (\ref{omegan}), is responsible for the increasingly asymmetrical forms of the resuscitations that we saw in Fig.~2. 

\subsubsection{Normalization}
In Appendix D we use our matched asymptotics solutions from Appendix C to determine the normalization factors $Z_n$, finding
\begin{align}\label{ZTfinal0}
   Z_n^2 &= \frac{d k_n}{dn}\;,
\end{align}
where we differentiate $k_n$ with respect to $n$ as if $n$ in (\ref{kn}) were a continuous variable.

This result is convenient because the factor $dk_n/dn$ will allow us to approximate sums over $n$ as integrals over $k$, at least in some ranges of $t$. As we will see, this mathematical convenience turns out to be what keeps the breathing of the trapped soliton looking just like the breathing of the untrapped soliton until times of the order of the trap period.

\section{IV. Resuscitations}
\subsection{IV.A The breathing mode amplitude $B(t)$}
\begin{figure}[!h]
    \centering
    \includegraphics[width=0.45\textwidth]{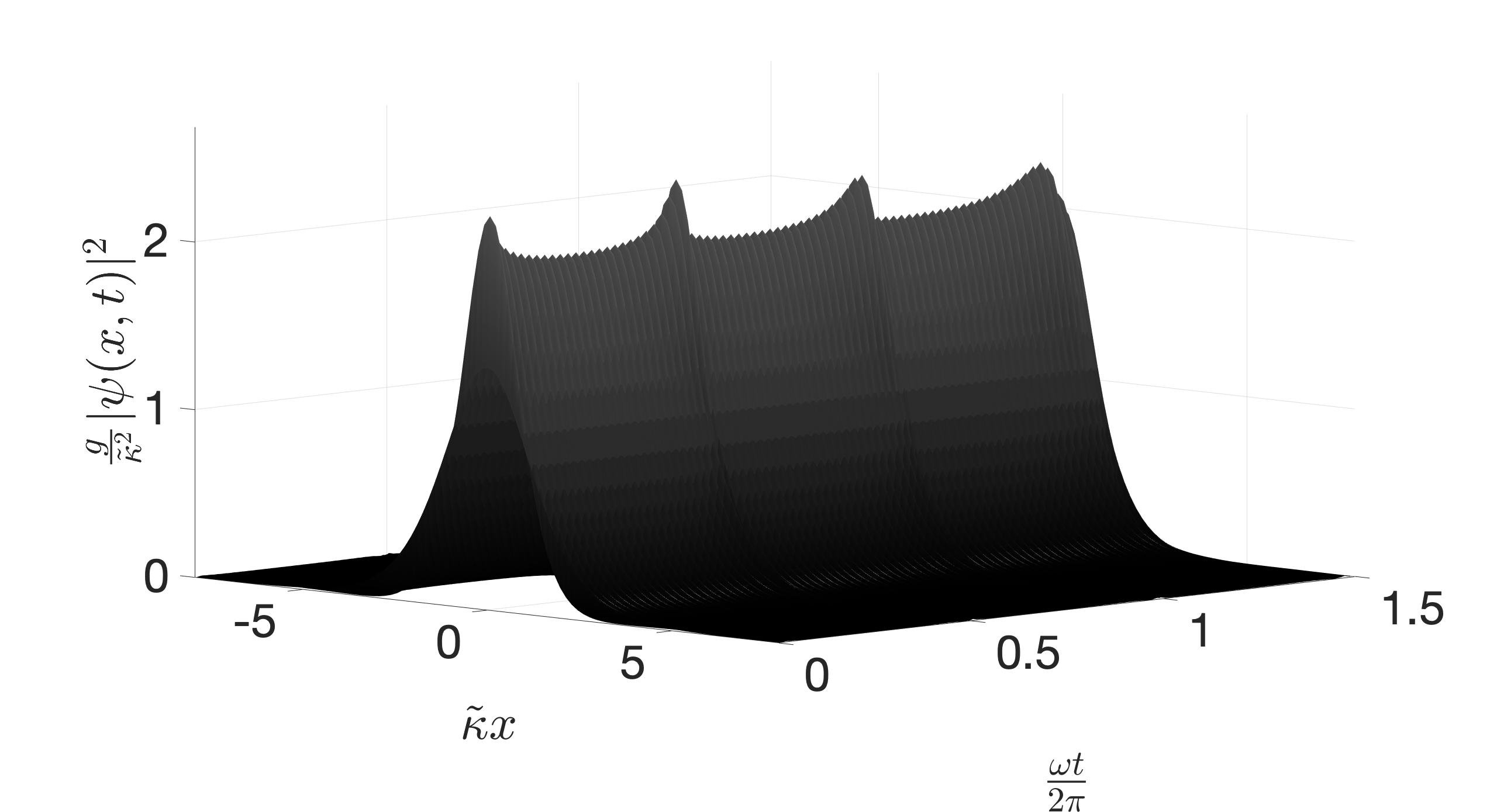}
    \caption{Condensate density in space and time, for an initially excited breathing mode in a harmonic trap with frequency $\omega$.}
\end{figure}
Figure 3 shows an example of the full density profile of the soliton breathing in a harmonic trap. Over many of its short periods, the mode is similar to that of the untrapped soliton, with a $1/\sqrt{t}$ dispersive decay in its amplitude. Over longer times, however---in particular after about one-half of the trap period---the amplitude increases again, showing periodic revivals, yet of an asymmetrical form. The behavior of the full density profile follows its peak at $x=0$ in a simple way, and so we focus on $|\psi(0,t)|$, which exhibits the trumpet-shaped asymmetrical resuscitations that we showed in Fig.~2. 

For our small initial perturbation (\ref{Psiex}) from a perfect bright soliton, we have
\begin{align}\label{Psiex0}
\psi(0,t) &= \Psi_0(0,t)+e^{i\frac{\hbar\kappa^2}{2M}t}\delta\psi(0,t)\nonumber\\
&= e^{i\frac{\hbar\kappa^2}{2M}t}\left(\frac{\kappa}{\sqrt{g}}+\delta\psi(0,t)\right)\nonumber\\
\delta\psi(0,t) &= \sum_{n=0}^\infty \left(\beta_n e^{-i\Omega_n t} u_n(0) + \beta_n^* e^{i\Omega_n t}v_n^*(0)\right)\nonumber\\
&=\frac{2Z_n}{\sqrt{2\pi}}\sum_{n=0}^\infty \frac{\beta_n e^{-i\Omega_n t} k_n^2 + \beta_n^* e^{i\Omega_n t}\kappa^2}{(k_n^2+\kappa^2)}\;,
\end{align}
when we use \begin{equation}(u_n,v_n)=Z_n[(u_{k_n},v_{k_n})+(-1)^n(u_{-k_n},v_{-k_n})]\end{equation} and insert $u_k(0),v_k(0)$ from (\ref{Kaup}). The discrete $\beta_n$ are defined from the initial $\delta\psi(x,0)$ by projection onto the complete set of discrete $(u_n,v_n)$:
\begin{equation}\label{betan}
    \beta_n = \int\!dx\, \left(u^*_n(x)\delta\psi(x,0) - v^*_n(x)\delta\psi^*(x,0)\right)\;.
\end{equation}
With $\delta\psi(x,0)$ still given by our initial breathing perturbation (\ref{deltabr}), all $\beta_n$ with odd $n$ vanish by parity. Furthermore $\delta\psi(x,0)$ essentially vanishes outside the Soliton Zone, and so the integral over $x$ in (\ref{betan}) is the same as the analogous integral for $\beta_k$ in the trap-free case (\ref{betak1}), just with $k\to k_n$ and an additional factor of $2Z_n$:
\begin{equation}
    \beta_{n} =  -\sqrt{\frac{2\pi}{g}}\,\delta\,Z_n\mathrm{sech}\Big(\frac{\pi k_{n}}{2\kappa}\Big)
\end{equation}
for all even $n$, but $\beta_n = 0$ for all odd $n$. Since $\beta_n$ are thus real, we obtain a compact sum over even $n$ for $|\psi(0,t)|^2$ to order $\delta$:
\begin{align}\label{summm}
|\psi(0,t)|^2 &= \frac{\kappa^2}{g}[1 - \delta\,B(t) +\mathcal{O}(\delta^2)]\nonumber\\
B(t) &= \frac{4}{\kappa}\sum_{n=0}^\infty Z_{2n}^2 \mathrm{sech}\Big(\frac{\pi k_{2n}}{2\kappa}\Big)\cos(\Omega_{2n}t)\;.
\end{align}
We identify this $B(t)$ as the trapped soliton's breathing amplitude, analogous to the $B_0(t)$ for the untrapped soliton in (\ref{breathe}).

\subsubsection{Evaluating $B(t)$ for small $t\ll \omega^{-1}$}
To evaluate $B(t)$ explicitly, we combine our results (\ref{omegan},\ref{kn},\ref{deltan3},\ref{ZTfinal0}) for $\Omega_n$, $k_n$, and $Z_n$, in the form\begin{widetext}
\begin{align}\label{Bdef}
    B(t) &= 2\mathrm{Re}\left(e^{i\frac{\hbar\kappa^2}{2M}t}e^{i\frac{\omega}{2}t}\sum_{n=0}^\infty \frac{1}{\kappa}\frac{dk_{2n}}{dn} \mathrm{sech}\Big(\frac{\pi k_{2n}}{2\kappa}\Big)e^{2i\omega t\left(n+\frac{1}{\pi}\cos^{-1}\frac{1-2\varepsilon^2n}{1+2\varepsilon^2n}\right)}\right)\nonumber\\
    &=2\mathrm{Re}\left(e^{i\frac{\hbar\kappa^2}{2M}t}\sum_{n=0}^\infty \frac{1}{\kappa}\frac{dk_{2n}}{dn} \mathrm{sech}\Big(\frac{\pi k_{2n}}{2\kappa}\Big)e^{i\frac{\hbar k_{2n}^2}{2M}t}\right)\;.
\end{align}\end{widetext}
As long as each term in the sum varies slowly with $n$, the sum approaches a Riemann sum and we can use the Euler-Maclaurin series 
\begin{align}\label{int_app}
    \sum_{n=0}^\infty f(\Delta n)=\frac{1}{\Delta}\int_0^\infty dx f(x)-\left.\frac{f(x)}{2}\right|_0^\infty+\mathcal{O}(\Delta^2)
\end{align}
to approximate (\ref{Bdef}) as the $B_0(t)$ (\ref{breathe}) that we found for the breathing amplitude of the untrapped soliton:
\begin{equation}\label{int2}
    B(t)= 2\mathrm{Re}\left(e^{i\frac{\hbar\kappa^2}{2M}t}\int_0^\infty\!\frac{dk}{\kappa}\, \mathrm{sech}\Big(\frac{\pi k}{2\kappa}\Big)e^{i\frac{\hbar k^2}{2M}t}\right)+\mathcal{O}(\varepsilon)\;,
\end{equation}
where the factor of two appears instead of extending the lower limit of integration to $-\infty$, as in (\ref{breathe}), since the integrand is even in $k$.

Do the terms in the sum (\ref{Bdef}) actually vary slowly with $n$, though? Whenever $n$ appears in the summand (\ref{Bdef}), it is indeed multiplied by the small factor $\varepsilon^2$---except for the one term $2i\omega t n$ that appears in the exponent. For times much less than the trap period, this term also varies slowly with $n$, and so for $t\ll 1/\omega$ the trapped bright soliton still breathes just as it does without any trap, and the breathing amplitude decays in a $1/t$ envelope. For times $t\gtrsim 1/\omega$, however, the sum (\ref{Bdef}) is \emph{not} well approximated by the integral (\ref{int2}). The non-Markovian nature of the Trap Zone as an environment for the breathing soliton becomes apparent. 

\subsection{IV.B Evaluating $B(t)$ for $t=N\pi/\omega + \Delta t$}
For general times of order $t\sim 1/\omega$, the sum (\ref{Bdef}) has no simple evaluation. We can consider times close to a whole number of half-periods, however, 
\begin{equation}\label{Deltatdef}
    t = \frac{N\pi}{\omega}+\Delta t
\end{equation}
for whole-number $N$ and $|\Delta t|\ll 1/\omega$. For $N\gtrsim \varepsilon^{-2}$, $2N\cos^{-1}\big((1-2\varepsilon^2 n)/(1+2\varepsilon^2 n)\big)$ does not vary slowly enough with $n$ for any integral approximation to be valid, but as long as $N\ll\varepsilon^{-2}$, the summand in (\ref{Bdef}) again varies slowly with $n$, because $e^{2i\pi Nn} = 1$. This yields the modified integral\begin{widetext}
\begin{align}
    B(t)&= \mathcal{O}(\varepsilon)+2 \mathrm{Re}\left(e^{i\frac{N\pi}{2}}e^{i\frac{\hbar\kappa^2}{2M}t}\int_0^\infty\!\frac{dk}{\kappa}\, \mathrm{sech}\Big(\frac{\pi k}{2\kappa}\Big)e^{i\frac{\hbar k^2}{2M}\Delta t}e^{2iN\cos^{-1}\big(\frac{\kappa^2-k^2}{k^2+\kappa^2}\big)}\right)=:B_N(\Delta t)+\mathcal{O}(\varepsilon)
    \end{align}
    for
\begin{align}\label{BNtau}    B_N(\Delta t) &=2\mathrm{Re}\left[e^{i N\pi\varepsilon^{-2}}e^{i\frac{N\pi}{2}}e^{i\frac{\hbar\kappa^2}{2M}\Delta t}\int_0^\infty\!dz\, \mathrm{sech}\Big(\frac{\pi z}{2}\Big)e^{i\frac{\hbar\kappa^2}{2M}\Delta t\, z^2}\left(\frac{1+iz}{1-iz}\right)^{2N}\right]\;,
\end{align}\end{widetext}
using $k\to\kappa z$ as the integration variable and noting that
\begin{equation}
    \cos^{-1}\left(\frac{1-z^2}{1+z^2}\right) \equiv i\ln\left(\frac{1-iz}{1+iz}\right)\;.
\end{equation}

After Eqns.~(\ref{kn}-\ref{deltan3}) for the collective mode spectrum, Eqn.~(\ref{BNtau}) for the resuscitation profiles in time is our paper's final major result. What does it tell us?

\subsection{IV.C Features of the resuscitation episodes $B_N(\Delta t)$}

Equation (\ref{BNtau}) defines the oscillation amplitude of the soliton's peak density for a range of time $\Delta t$ around the $N$th revival time, $t = N\pi/\omega + \Delta t$ as defined in (\ref{Deltatdef}). Even before we attempt to evaluate the integral over $z$ in (\ref{BNtau}), we can see that the two relevant frequency scales $\omega$ and $\hbar\kappa^2/(2M)$ appear in $B_N(\Delta t)$ in quite separate roles.

\subsubsection{The phase shift $N\pi/\varepsilon^2$}
First of all, the trap frequency $\omega$ only appears in (\ref{BNtau}) in one single place: it is inside the large factor
\begin{equation}
    \varepsilon^{-2} = \frac{\hbar\kappa^2}{2M\omega}
\end{equation}
which appears in the first phase in $B_N(\Delta t)$. In our small-$\varepsilon$ limit this phase is enormous for $N\geq 1$. 

This large phase must indeed really be present, since the number of breathing mode periods in one-half of a trap period is a large number that does not have to be an integer. It is not easy, however, to make a reliable prediction of the value of this huge phase modulo $2\pi$. All of our calculations have neglected higher orders of $\varepsilon$, and if any of these neglected corrections turn out to shift the overall phase of the breathing mode, their effects on the phase may well be as great as $N\pi\varepsilon^{-2}$ modulo $2\pi$, even if they are much smaller than $N\pi\varepsilon^{-2}$.

Even if we had been able to perform our Bogoliubov-de Gennes calculations to all orders in $\varepsilon$, furthermore, it would still be beyond the scope of this paper to assess possible additional secular phase drifts due to nonlinear mean-field effects, or quantum correction to mean-field physics, or limitations of the quasi-one-dimensional approximation, all of which could likewise shift the breathing mode phase over a large number of breathing periods. And, on the other hand, even if we could compute all these effects, it is unlikely that the soliton chemical potential $\hbar^2\kappa^2/(2M)$ can  experimentally be controlled precisely enough to make $N\pi\varepsilon^{-2}$ modulo $2\pi$ a reproducible phase instead of an effectively random one.

The phase factor $e^{iN\pi\varepsilon^{-2}}$ in (\ref{BNtau}) must probably therefore be considered uncertain, or even randomly varying in each experimental run. All the other features of $B_N(\Delta t)$---the frequency of the breathing oscillations, and their asymmetrically reviving and decaying amplitude envelope---should in contrast be robust predictions of mean field theory, as linearized for a small-amplitude collective excitation. 

\begin{figure*}[!t]
    \includegraphics[width=\textwidth]{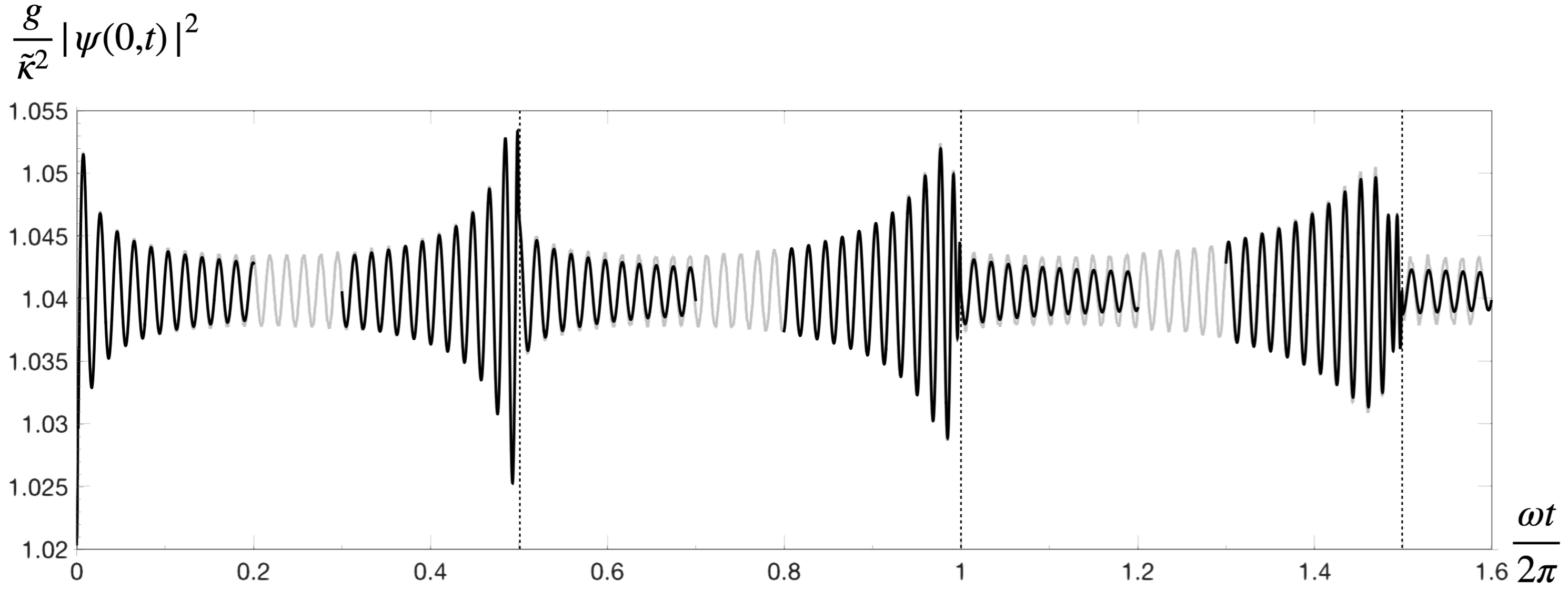} 
    \caption{\label{trumpfinal}The same resuscitation sequence shown in Fig.~\ref{trump} is here rendered faintly in the background, for the same $\delta = 10^{-2}$ and $\varepsilon = (1+\delta)^{-2}/\sqrt{50}\doteq 0.14$. Superimposed in four time intervals around the revival times $t=N\pi/\omega$ are the Bogoliubov-de Gennes approximations $(1+\delta)^2[1-\delta B_N(t-N\pi/\omega)]$, where the integral approximation (\ref{BNtau}) has been used for $B_N(\Delta t)$, with the integrals evaluated numerically.}
\end{figure*}

\subsubsection{Independence of $\omega$}
Unlike the doubtful $e^{iN\pi\varepsilon^{-2}}$ phase factor, the other features of $B_N(\Delta t)$ are all completely independent of the trap frequency $\omega$, as long as we are indeed in the $\omega\ll \hbar\kappa^2/M$ limit that we have always assumed in this paper. In all cases in this regime, the soliton's breathing oscillation always has the same frequency scale $\hbar\kappa^2/(2M)$ as the untrapped soliton. As we will see, the amplitude envelope around each resuscitation episode has features on the additional time scales $\hbar\kappa^2/(MN)$ and  $\hbar\kappa^2/(MN^2)$---but neither of these additional time scales involves the trap frequency $\omega$. 

If we compare soliton breathing in traps of different $\omega$, therefore, we will have to wait different lengths of time between resuscitations, because the time between resuscitations is half the trap period, $\pi/\omega$. Once the resuscitation episodes begin in each trap, however, they will look the same in both cases, in spite of the different $\omega$. In this sense $B_N(\Delta t)$ should be considered independent of $\omega$, and universal for all $(2M\omega)/(\hbar\kappa^2)=\varepsilon^2\ll 1$ (once we interpret the huge phase $e^{iN\pi\varepsilon^{-2}}$ as effectively random modulo $2\pi$).

\subsubsection{Form of $B_N(\Delta t)$ for smaller $N$}
For $N=0$, $B_0(\Delta t)=B_0(-\Delta t)$, since $\Delta t \to-\Delta t$ just changes the term in square brackets in (\ref{BNtau}) to its complex conjugate and this does not affect its real part. For $N>0$, however, this no longer true, and $B_{N>0}(\Delta t)$ is asymmetric in $\Delta t$, with the asymmetry becoming more marked with increasing $N$. Figure \ref{trumpfinal} shows a direct comparison between $g|\psi(0,t)|^2/\tilde{\kappa}^2$ from Fig.~\ref{trump} and the Bogoliubov-de Gennes approximations $(1+\delta)^2[1-\delta B_N(t-N\pi/\omega)]$ in intervals around three successive resuscitation times $t=N\pi/\omega$. Since $\varepsilon\doteq 0.14$ in this case is not actually very small, the integral approximation becomes poorer at larger $\Delta t$, and for later resuscitations it evidently exaggerates the severity of the post-revival drop in breathing amplitude. 

The agreement between nonlinear mean field theory and the Bogoliubov-de Gennes approximation is nevertheless close enough to confirm that the asymmetric profile of the resuscitations is indeed due to the slight non-uniformity of the collective mode frequency spectrum. Somewhat surprisingly, even the suspect phase $e^{iN\pi\varepsilon^{-2}}$ in $B_N(\Delta t)$ matches the nonlinear evolution very well; perhaps it helps that in this case $\varepsilon^{-2}\doteq 52$ happens to be very close to an even number.

The plots of $B_N(\Delta t)$ for $N\leq 3$ shown in Fig.~4 represent numerical evaluation of the $z$ integral in (\ref{BNtau}). To see what happens for larger $N$, we can pursue analytical approximations to (\ref{BNtau}) instead.

\subsubsection{Evaluating $B_N(\Delta t)$ for $1\ll N\ll \varepsilon^{-2}$}

While our figures have shown no more than three revivals, we can see that the asymmetry persists over many revivals, by applying the saddlepoint approximation to the integral over $z$ in (\ref{BNtau}), for large $\hbar\kappa^2 t/M$ and large $N$ (as long as $N\ll\varepsilon^{-2}$ so that the integral approximation (\ref{BNtau}) to the sum (\ref{Bdef}) remains good). The saddlepoint condition is the cubic equation
\begin{equation}
    z_s^3 + z_s + \frac{4NM}{\hbar\kappa^2 \Delta t}=0\;.
\end{equation}

For $\Delta t>0$ there are no saddlepoints $z_s$ near the positive real axis of $z$ which is our integration range, and the integral is dominated by its $z=0$ endpoint. For small $z$ and large $N$, the factor $((1+iz)/(1-iz))^{2N}\doteq e^{4iNz}$, and the $z$ integral in (\ref{BNtau}) becomes close to $i/(4N)$, leaving 
\begin{equation}\label{flatsine}
    B_N(\Delta) \xrightarrow[\Delta t>0]{} \frac{1}{2N}\cos\Big(\frac{N+1}{2}\pi + \frac{N\pi}{\varepsilon^2}+\frac{\hbar\kappa^2}{2M}t\Big)\;.
\end{equation}
Thus, immediately after every $N$th trap half-period, the breathing amplitude drops to $\delta/(2N)$---and then remains \emph{constant} for $\mathcal{O}(16N^2)$ breathing periods, before eventually beginning to attenuate as $|\Delta t|^{-1/2}$ (if $\varepsilon^2$ is small enough that the next resuscitation has not yet begun by this point). This simple behavior for $\Delta t>0$ appears at leading order in $1/(4N)$, and so we can already see it beginning in the right-most plot in Fig.~4, at just $N=3$.

For $\Delta t<0$, in contrast, the $z$ integrand in (\ref{BNtau}) has a saddlepoint $z_s$ on the positive real axis of $z$. The location of the saddlepoint depends on both $\Delta t$ and $N$, so that the oscillation amplitude $B_N(\Delta t)$ has a non-trivial envelope for negative $\Delta t$, unlike the constant amplitude $\propto 1/N$ for positive $\Delta t$. The saddlepoint approximation is remarkably close to the numerical $B_N(\Delta t)$ as soon as $-\hbar\kappa^2\Delta t/M\gg 1$, but it is a complicated expression unless $-\hbar\kappa^2 \Delta t/M \gg N$, so that the cubic solution for $z_s$ reaches the simple limit
\begin{equation}
    z_s \longrightarrow - \frac{4NM}{\hbar\kappa^2\Delta t}\ll 1\;.
\end{equation}

In this small-$z_s$ but large-$\hbar\kappa^2 t/M$ limit, the $z$ integral in (\ref{BNtau}) simplifies and we obtain\begin{widetext}
\begin{equation}
    B_N(\Delta t)\xrightarrow[\Delta t<0]{} \sqrt{\frac{\pi M}{-\hbar^2\kappa^2\Delta t}}\mathrm{Re}\left(e^{i N \pi \varepsilon^{-2}}e^{iN\frac{\pi}{2}}e^{i\frac{\hbar\kappa^2\Delta t}{2M}}e^{-8iN^2\frac{M}{\hbar\kappa^2 \Delta t}}\Big[1+\mathrm{erf}\Big((1+i)\sqrt{\frac{4N^2M}{-\hbar\kappa^2 \Delta t}}\Big)\Big]\right)\;,
\end{equation}\end{widetext}
where $\mathrm{erf}(z)$ is the error function.
In stark contrast to the result (\ref{flatsine}) for $\Delta t>0$, for $\Delta t<0$ the breathing amplitude $B_N(\Delta t)$ has a rather complicated modulating envelope that can extend over very many breathing periods. The dramatic asymmetry of the resuscitations of trapped soliton breathing modes can thus indeed persist over many revivals.

\section{V. Discussion}

\subsection{V.A Revivals}
The initial attenuation of the soliton breathing mode within a $1/\sqrt{t}$ envelope is already expected from the case with no trapping potential. In the trapped case, it is then only natural to expect revivals after every trap half-period, once we realize that the attenuation of the excitation involves emission of particles, since a half-period is how long it takes before any initially emitted atoms will begin to return to the soliton. The revivals occur every half-trap-period, which is easy to deduce because our initial perturbation of the soliton is spatially even. In a trap, these even collective modes have eigenfrequencies $\Omega_{2n}$ separated by approximately $2\omega$, leading to a revival time estimate of $\pi/\omega$.

The forms of the envelopes of the breathing amplitude at successive resuscitations are the more surprising phenomenon that we have seen in this paper. These breathing revivals have an asymmetric profile, which becomes more pronounced in later resuscitations. The breathing amplitude actually starts to revive distinctly \emph{earlier} than the first $\pi/\omega$ half-period, and this early revival becomes steadily stronger in later resuscitations. Almost immediately after the half-period, then, the breathing amplitude falls sharply, before more gradually decaying until the next resuscitation.

\subsection{V.B A qualitative explanation}
Qualitatively, we can explain the asymmetrical resuscitation phenomenon simply, from the spectral facts that we have described above. The two lowest modes of the Bogoliubov-de Gennes spectrum remain as the time-translation zero mode and the collective dipole mode with eigenfrequency exactly $\omega$, while all the other modes are raised above a large gap $\hbar\kappa^2/(2M)$, because excitations out of the condensate must all be excited above the large mean-field binding energy of the soliton. When the soliton is much narrower than the trap width, the spectrum above this chemical potential gap is still approximately like that of a single harmonic oscillator. Nevertheless the entire above-gap spectrum has to be slightly distorted away from the even level spacing of an oscillator, to ``replace'' the missing two lowest modes. These distortions are indeed supplied by the differences between the Bogoliubov-de Gennes equations and the Schr\"{o}dinger equation of a single trapped particle.

In particular the slight systematic spectral distortion is such that the gaps between successive even-mode eigenfrequencies are consistently a bit \emph{greater} than $2\omega$. This means that quasi-particles oscillate back and forth in the trap at slightly \emph{higher} frequencies than $\omega$. Why are the atoms sped up in this way?

The atoms cannot oscillate faster than $\omega$ while they are in the Trap zone, outside the soliton's mean field. While they are inside the soliton, however, the atoms evidently experience an effectively negative potential, giving them additional kinetic energy: higher speed. Although the soliton occupies only a narrow region in the middle of the trap, within this narrow region quasiparticles effectively experience quite a deep negative potential, and speed up significantly---as long as they are inside the soliton.

This effective burst of speed as quasiparticles pass through the soliton background lets them systematically return to the soliton again slightly earlier than they would in the trap with no soliton. Since the spectral stretching $d\nu/dn$ is not uniform, but affects lower eigenfrequencies more than higher ones in accordance with (\ref{deltan3}), there is a range of early return times. With every subsequent resuscitation, moreover, the lead of those faster quasiparticles steadily increases, in comparison to where they would be after the same amount of time without the speed boosts while they are inside the soliton. They return earlier and earlier every time, so that subsequent resuscitations start earlier and earlier.

The particles that return to the soliton before a full half-period keep moving on through the soliton, moreover, to continue oscillating back and forth in the trap. Shortly after the half-period, the early arrivals are already gone again, and the early amplification that they contributed to the revived breathing amplitude is now missing. This explains why the breathing amplitude drops after each half-period, and drops more and more noticeably in later resuscitations.

\subsection{V.C The effect is not trivial}
The asymmetric resuscitation profile should thus be comprehensible, given the spectral stretching that we have described in terms of speed boosts while within the soliton's attraction. To improve upon this qualitative explanation, however, we have not been able to find any satisfactory analysis that is any simpler than the full Bogoliubov-de Gennes calculation that we have explained in our text and Appendices. The spectral distortion, and speeding-up that it implies for quasiparticles, is a non-trivial effect of coherent interaction between excited atoms and the background condensate. 

The breathing excitations that we describe consist mostly of low-frequency quasiparticles (that is, with $\Omega_n$ close to the $\hbar\kappa^2/(2M)$ top of the mean-field gap). Outside the narrow background soliton, these quasiparticles are simply atoms, moving in a harmonic potential. Within the soliton background, however, these excited atoms are dressed by many-body effects: they become quasiparticles that involve many atoms. As quantum many-body effects go, these effects are still simple, inasmuch as they can be described within the linearized mean-field theory of Bogoliubov-de Gennes. Simpler than this, though, they are not.

The breathing mode of a bright soliton shakes particles out of the soliton. This means that a slightly imperfect bright soliton, with a breathing excitation, is an intrinsic atomic antenna, emitting atoms coherently with an inherent frequency that is defined, up to the limit of dispersive broadening, by the atomic interactions themselves. If the soliton breathes in a trap, the coherently emitted atoms return periodically. In what is perhaps one of the simplest possible examples of back-reaction, they interfere with their emitter itself. A bright soliton breathing in a trap is an intrinsic atomic interferometer.

Mean-field physics of dilute Bose-condensed gases is in general well understood, and so it is more complicated quantum many-body effects that may be of greater physical interest in experiments with trapped solitons. As a distinctive and non-trivial mean-field prediction, then, the asymmetrical resuscitations of the soliton breathing mode may be a useful background against which to look for subtler quantum effects, such as decay of the resuscitations due to decoherence from interactions among quasi-particles.

\subsection{V.D Implications for non-Markovian evolution}
Non-Markovian environments are in general so much more difficult to treat than Markovian ones that non-Markovian effects are often ignored, because even if they are not really small, they are often small in comparison to the effort required to compute them. If the open-system dynamics are represented in an equation of motion for system degrees of freedom alone, the memory effect of a non-Markovian environment typically makes this equation of motion integro-differential rather than differential. Integral equations of almost any kind are much harder to solve than purely differential equations.

Explicit results for non-Markovian open systems can sometimes be obtained by solving the evolution of the larger closed system that includes the environment---especially when this larger-system dynamics can be treated as linear, so that the non-Markovian evolution can be obtained by projection from a linear evolution of a large number of degrees of freedom. The present paper has been an example of this approach. And it has shown how even as simple a form of environment memory as oscillation in a weak harmonic trap can produce strange effects like our asymmetric resuscitations, when the simply non-Markovian environment interacts with a non-trivial open system such as the breathing mode of an atomic bright soliton.

\subsection{V.E Analogous effects in higher dimensions?}

The small parameter $\varepsilon = 1/(\kappa a_0)$, the ratio of soliton width to trap width, has been important in enabling all our calculations. It is also essential conceptually, however, for our view of the soliton breathing mode as an open quantum system in a non-Markovian environment. If the soliton width and the trap width were comparable, we would simply have a trapped condensate with an even collective mode; there would be no basis for the open-system picture, and the multiple time scales of breathing, attenuation, and revival would not exist.

The length scale hierarchy $\kappa^{-1}\ll a_0$, and the associated time scale hierarchy $\omega\ll \hbar\kappa^2/(2M)$, exist in this quasi-one-dimensional scenario only because in one dimension a bright soliton defines its own width, through the attractive nonlinearity, independently of any trapping confinement. This kind of self-confinement independently from a trap does not occur in more than one dimension, but one might consider radial breathing of a vortex core as kind of inverted analog of our scenario, in which it is a small void within a much larger condensate that breathes, instead of a small condensate in large empty trap.

Collective excitations of vortex cores in quasi-two-dimensional condensates with attractive interactions have been studied in \cite{attvortex}, where in particular a regime of large-amplitude collective excitations is found in which a condensate with a vortex repeatedly breaks up into two clouds that then recombine. While our breathing mode with emission of atoms can be considered a smaller-amplitude version of this behavior, the fact that in the 2D scenario the whole condensate, the vortex core, and the trap width are all of comparable scale prohibits the kind of multi-scale evolution that can be seen as short-period breathing with attenuation and revivals over a much longer period. The prospects for higher-dimensional analogs to our resuscitations seem to be inherently limited for attractively interacting condensates.

In repulsively interacting condensates, on the other hand, a large condensate can have a vortex with a much smaller core, since the core size is set by the interacting strength and the local condensate density. A breathing mode of a vortex core in a repulsively interacting condensate has been studied in \cite{repvortex}, with frequencies found to be on the order of the chemical potential scale, as it is in our soliton case. The time-dependent variational Ansatz approach used in \cite{repvortex} assumes that core radius oscillation is a single collective mode, however, and so does not allow for the possibility that it is a component in a quasi-continuum of many modes and therefore decays through dispersion. Hence it is not clear from \cite{repvortex} whether vortex core breathing might attenuate and then revive as our soliton breathing mode does. It may in future be worth performing a similar analysis to the one that we have offered in this paper, based on identifying breathing as a component in a complete set of collective modes, for vortex cores in large, repulsively interacting condensates in two or three dimensions.

\section*{Acknowledgement}

This project traces its origins to T. Schlachter’s keen observation and early encouragement. We gratefully acknowledge his role in inspiring its initial direction.

\section{Appendix A: Quantum depletion is one-third of an atom}
\subsection{A.1: Quantum fluctuations}
The Gross-Pitaevskii nonlinear Schr\"odinger equation (\ref{alpha_2}) is a classical approximation to the Heisenberg equation of motion for the second-quantized atom destruction field operator $\hat{\psi}(x,t)$,
\begin{equation}\label{HEM}
    i\frac{\partial}{\partial t}\hat{\psi} = -\frac{\hbar}{2M}\frac{\partial^2}{\partial x^2}\hat{\psi}+V(x)\hat{\psi} - \frac{\hbar g}{M}\hat{\psi}^\dagger\hat{\psi}^2\,
\end{equation}
when the actual two-body interaction potential is approximated with a Fermi-Huang pseudo-potential having one-dimensional attractive scattering strength $g$. To go beyond the mean field approximation, we can define a new second-quantized destruction field $\delta\hat{\psi}(x,t)$ by subtracting the mean field $\Psi_0(x,t)$ from the atom destruction field:
\begin{equation}
    \hat{\psi}(x,t) = \Psi_0(x,t)+e^{i\frac{\hbar\kappa^2}{2M}t}\delta\hat{\psi}(x,t)\;.
\end{equation}
The classical limit which produces the c-number nonlinear Schr\"odinger equation (\ref{alpha_2}) is then simply $N_0\to\infty$ for fixed $\kappa$, so that $\Psi_0$ is of order $N_0^{1/2}$ while $\delta\hat{\psi}$ remains of order $N_0^0$ and can therefore be neglected in comparison to $\Psi_0$.

(To make it explicit in our bright soliton case that the background mean field $\Psi_0$ is proportional in amplitude to $\sqrt{N_0}$, we can express (\ref{alpha_1}) equivalently as
\begin{equation}
    \Psi_0(x,t) = e^{i\frac{\hbar\kappa^2}{2M}t}\sqrt{\frac{N_0}{2}}\sqrt{\kappa}\,\mathrm{sech}(\kappa x)\;,
\end{equation}
using the relation (\ref{number}) between $g$, $\kappa$, and $N_0$.)

We can then confirm by integration that we preserve the canonical commutation relations
\begin{align}
    [\hat{\psi}(x,t),\hat{\psi}^\dagger(x',t)] &= [\delta\hat{\psi}(x,t),\delta\hat{\psi}^\dagger(x',t)] &= \delta(x-x')\nonumber\\
    [\delta\hat{\psi}(x,t),\delta\hat{\psi}(x',t)]&=0
\end{align}
and also solve the Heisenberg equation of motion (\ref{HEM}) to linear order in $\delta\hat{\psi}$ with the expansion in $u_k(x)$ and $v_k(x)$ modes given in (\ref{BdG}) of the main text, simply replacing the c-number coefficients $\beta_k$ with canonical bosonic destruction operators $\hat{b}_k$. (We also quantize the zero mode co-efficients as position and momentum operators, $q_\pm\to\hat{q}_\pm$ and $p_\pm\to\hat{p}_\pm$, satisfying the canonical commutation relation.) This yields
\begin{align}\label{BdGq}
    \delta\hat{\psi}(x,t) =& \sum_\pm\Big[ (\hat{q}_\pm + \hat{p}_\pm t)R_\pm(x) + i\hat{p}_\pm S_\pm(x)\Big]\nonumber\\
    &+\int_{-\infty}^\infty\!dk\,\Big[\hat{b}_k e^{-i\Omega(k)t}u_k(x) + \hat{b}^\dagger_k e^{i\Omega(k)t}v^*_k(x)\Big]\nonumber\\
    \Omega(k) =& \frac{\hbar}{2M}(k^2+\kappa^2)
\end{align}
as the quantum version of (\ref{BdG}).

\subsection{A.2: Universally small quantum depletion}

The \textit{quantum depletion} is a benchmark for the mean field approximation of assuming that all bosons are occupying the one orbital with wave function proportional to $\Psi$. The depletion is just the expected number of atoms that are not in the condensate, when the system is in the Bogoliubov-de Gennes ground state, namely the quasiparticle vacuum state which is annihilated by all $\hat{b}_k$. The two discrete Bogoliubov-de Gennes modes do have quantum fluctuations, but these are not normally seen as representing any displacement of atoms out of the mean field orbital; they are rather considered as possible quantum uncertainties in the total atom number and in the position and momentum of the center of mass. Ignoring these discrete modes, therefore, the expectation value of the gas density in the quasiparticle vacuum state is then
\begin{align}\label{depletion}
    \rho(x) &= \langle\hat{\psi}^\dagger\hat{\psi}\rangle = |\Psi_0|^2 + \int\!dk\,|v_k|^2\nonumber\\
    &=N_0\frac{\kappa}{2}\,\mathrm{sech}^2(\kappa x) + \frac{\kappa}{4}\mathrm{sech}^4(\kappa x)\;.
\end{align}

Eqn.~(\ref{depletion}) implies the presence of some non-condensed atoms---atoms that are not occupying the $N_0^{-1/2}\Psi_0$ orbital but are instead more tightly concentrated inside the soliton core. We can understand this \emph{quantum depletion} of the condensate by noting that an ideal Bose-Einstein condensate, with all particles occupying the same orbital, has no correlations between particles. Scattering between particles introduces at least pair correlations, however, and in second-quantized terms these imply population in additional orbitals. For a bright soliton, however, the total quantum depletion 
\begin{align}
    \Delta N =& \int\!dx\,\langle\hat{\psi}^\dagger\hat{\psi}\rangle - N_0
    = \int\!dx\,\frac{\kappa}{4}\mathrm{sech}^4(\kappa x)=\frac{1}{3}
\end{align}
turns out to be a universal number for all bright solitons, independent of both $g$ and $N_0$. It is exactly 1/3: not one-third of the gas, but \emph{one-third of an atom}. 

Since the number of condensate atoms $N_0$ can in principle be arbitrarily large, and in experiments can be thousands, this universally small quantum depletion suggests that the Gross-Pitaevskii mean field theory should be excellent, at least as long as our mean field does not depart too much from the perfect soliton profile $\Psi_0$. The problem of infra-red divergent quantum depletion, which one encounters in quasi-one-dimensional Bose condensates with repulsive interactions, does not occur for the attractively interacting bright soliton, because the self-limiting spatial extent of the soliton provides an intrinsic infra-red cutoff.

\section{Appendix B: Bogoliubov-de Gennes perturbations}
\subsection{B.1: Zero modes}

Small time-dependent perturbations around the untrapped bright soliton background can be expanded in a complete set of normal modes including two zero modes:
\begin{align}
    \delta{\psi}(x,t) =& \sum_\pm\Big[ {q}_\pm(t) R_\pm(x) + i{p}_\pm(t)S_\pm(x)\Big]\nonumber\\
    &+\int_{-\infty}^\infty\!dk\,\Big[{\beta}_k e^{-i\Omega(k)t}u_k(x) + {\beta}^*_k e^{i\Omega(k)t}v^*_k(x)\Big]\nonumber\\
    \Omega(k) =& \frac{\hbar}{2M}(k^2+\kappa^2)
\end{align}
for arbitrary (small) complex constants $\beta_k$ as well as real $q_\pm$, $p_\pm$.
The normalized mode functions $u_k(x),v_k(x)$ as found by \cite{Kaup} are shown in Eqn.~(\ref{Kaup}) of our main text. The two additional discrete modes $(R_\pm,S_\pm)$ are
\begin{align}\label{RS}
    \left(\begin{matrix} R_+(x)\\ S_+(x)\end{matrix}\right) &=\left(\begin{matrix} i\frac{\kappa^{3/2}}{\sqrt{2M}}\mathrm{sech}(\kappa x)\\ \frac{-i\sqrt{M}}{\sqrt{2\kappa}\hbar}\left(1-\kappa x\,\!\tanh(\kappa x)\right)\mathrm{sech}(\kappa x)\end{matrix}\right)\nonumber\\
    \left(\begin{matrix} R_-(x)\\ S_-(x)\end{matrix}\right) &=\left(\begin{matrix} \frac{\kappa ^{3/2}}{\sqrt{2M}}\mathrm{sech}(\kappa x)\tanh(\kappa x)\\ \frac{\sqrt{M\kappa}}{\hbar}\;x\,\mathrm{sech}(\kappa x)\end{matrix}\right)\;.
\end{align}
Note that $(R_+,S_+)$ are even functions of $x$, while $(R_-,S_-)$ are odd functions.

From the forms of these $R_\pm(x)$ and $S_\pm(x)$ it is possible to recognize that the discrete modes $({q}_\pm,{p}_\pm)$ are perturbations to the condensate global phase and number $N_0$, and to the  position and velocity of the soliton's center of mass, respectively. With $V(x)=0$ the center of mass velocity $p_-$ remains constant in time, while the center of mass position correspondingly moves at constant speed, $q_-(t)=q_-(0)+p_- t$. When the soliton is held in a harmonic trap $V(x) = M\omega^2 x^2/2$, the center of mass mode of the soliton becomes a harmonic oscillator at the trap frequency $\omega$ \cite{Kohn}. The harmonic discrete mode $(q_-,p_-)$ will not concern us in this paper, however, because we will analyse only the particular kind of small perturbations around $\Psi_0$ that evolve from slightly stretched or compressed initial conditions with even parity in $x$, so that $q_- = p_- = 0$ and $\beta_{-k}=\beta_k$.

The even discrete mode $(q_+,p_+)$, in contrast, is not constrained to have zero amplitude by parity. It must nonetheless also have zero amplitude, for the different reason that it is a \emph{zero mode} which can and must be ``fixed''.

\subsection{B.2: Freedom to fix the zero mode}
To see first of all that we always \emph{can} set $q_+$ and $p_+$ to zero, we can note that
\begin{align}
    \Psi_0\, +\, & q_+ R_+ + ip_+ S_+ \nonumber\\
    &= \frac{\kappa + \delta\kappa}{\sqrt{g}}\mathrm{sech}[(\kappa+\delta\kappa)x]e^{i\delta\theta} +\mathcal{O}(\delta\kappa,\delta\theta)^2\nonumber\\
    \delta\kappa &= \sqrt{\frac{Mg}{2\kappa}}\frac{p_+}{\hbar}\nonumber\\
    \delta\theta &= \sqrt{\frac{g\kappa}{2M}}q_+\;.
\end{align}
Changing $q_+$ is thus equivalent to multiplying $\Psi_0$ by a constant phase $\delta\theta$, and changing $p_+$ is equivalent to adjusting $\kappa$---and thus $N_0$---for fixed $g$. 

This means that adjusting $q_+$ and $p_+$ is simply perturbing one bright soliton \emph{into another bright soliton} with a slightly different phase and atom number. Alternatively, therefore, we can absorb the trivial phase rotation $q_+$ by multiplying $\psi\to e^{i\theta}\psi$ for an arbitrary constant phase $\theta$, and absorb $p_+$ into $\kappa$, relabeling $\kappa+\delta\kappa\to\kappa$. By this procedure we can always choose to make $q_+=p_+=0$ if we wish, just by adjusting which bright soliton we take as our background $\Psi_0$.

\subsection{B.3: Necessity of setting $q_+$ and $p_+$ to zero}
In fact setting $p_+\to 0$ in particular is not just a convenient option that we may take if we choose. In our time-dependent solutions (\ref{BdG}), the $\beta_k$ are all multiplied by harmonic factors $e^{-i\Omega(k)t}$, but in contrast $p_+(t)=p_+(0)$ is constant just like $p_-$ for $V(x)=0$, while $q_+(t) = q_+(0)+p_+(0)t$ in general also grows linearly.
This means that, unless $p_+=0$, the $R_+(x)$ component of $\delta\psi(x,t)$ will grow steadily larger, until the whole Bogliubov-de Gennes linear approximation breaks down. We can avoid that disaster, and keep our approximation accurate for a much longer time, if we exercise our freedom to set $p_+\to0$ by adjusting the value that we identify as our background $\kappa$. And indeed, if we want to keep our approximation accurate over finite times, then we \emph{must} fix the $p_+$ zero mode in this way.

Since
\begin{align}
    \int\!dx\,[R_+^*u_k -R_+ v_k ] &= 0\nonumber\\
    \int\!dx\,[R_+^*S_- + R_+ S_-^*] &=0 =\int\!dx\,[R_+^*R_- + R_+ R_-^*] \nonumber\\
    \int\!dx\,[R_+^* S_+ + R_+ S_+^*]&= -\hbar^{-1}\;,
\end{align}
setting $p_+\to0$ means ensuring that
\begin{align}
    0&=i\frac{\sqrt{N_0 M}}{\kappa}\int\!dx\,[R_+^*(x)\delta\psi(x,0) - R_+(x)\delta\psi^*(x,0)]\nonumber\\
    &= \int\!dx\,[\Psi_0(x,0)\delta\psi(x,0) +\Psi_0^*(x,0)\delta\psi^*(x,0)]\nonumber\\
    &=\int\!dx\,\Big(|\psi(x,0)|^2 - |\Psi_0(x,0)|^2\Big) + \mathcal{O}(\delta\psi)^2\;.
\end{align}

In other words, for an initial $\psi(x,0)$ which is close to a bright soliton, we must identify our background $\Psi_0$ as the particular bright soliton to which $\psi(x,0)$ is closest, by setting
\begin{equation}
    \kappa = \frac{g}{2}\int\!dx\,|\psi(x,0)|^2
\end{equation}
in accordance with (\ref{number}). For example, for initial $\psi$ of the form
\begin{equation} \label{hotfix1}
    \psi(x,0) = (1+\delta)\frac{\tilde{\kappa}}{\sqrt{g}}\,\mathrm{sech}({\tilde{\kappa}x})
\end{equation}
for small $\delta$, we must set 
\begin{equation}
    \kappa = (1+\delta)^2\tilde{\kappa}
\end{equation}
as stated in (\ref{kappafix}) of our main text.

\section{Appendix C: Matched asymptotics}
We solve the Bogoliubov-de Gennes problem with the harmonic $V(x)$ in the limit of small $\varepsilon$, using the method of matched asymptotics. We consider a length $L\gg 1/\kappa$, of which the precise value is arbitrary, but we specify that although $L$ is much larger than the soliton width $1/\kappa$, $L$ is also simultaneously much \emph{smaller} than the trap width $a_0$:
\begin{equation}
    L \sim \sqrt{\frac{a_0}{\kappa}}=\varepsilon^{1/2}a_0 = \varepsilon^{-1/2}\frac{1}{\kappa}
\end{equation}
for some $\varepsilon\ll 1$.
This order of magnitude for the length $L$ ensures, first of all, that
\begin{equation}
    \mathrm{sech}(\kappa L) = \mathcal{O}\left(e^{-\frac{1}{\sqrt{\varepsilon}}}\right)
\end{equation}
is utterly negligible, so that any expressions $\propto \mathrm{sech}(\kappa x)$ can be neglected for $|x|\gtrsim L$. We can similarly set $\tanh(\kappa x)\to\mathrm{sgn}(x)$ for any $|x|\gtrsim L$ as well.
On the other hand, we will simultaneously have
\begin{equation}
    \sqrt{\frac{2M\omega}{\hbar}}L = \mathcal{O}(\sqrt{\varepsilon})\;,
\end{equation}
which will allow us to make small-argument approximations for functions of the form
\begin{equation}
f\Big(\sqrt{\frac{2M\omega}{\hbar}}x\Big)\;,
\end{equation}
whenever $|x|\lesssim L$.

\begin{widetext}And so we solve the linearized Gross-Pitaevskii equation for $\delta\psi(x,t)$ in the harmonic trap,
\begin{align}\label{BdGV0}
    i\frac{\partial}{\partial t}\delta\psi = -\frac{\hbar}{2M}\left[\left(\frac{\partial^2}{\partial x^2}+\kappa^2 -\Big(\frac{M\omega}{\hbar}\Big)^2x^2 + 2\kappa^2 \mathrm{sech}^2(\kappa x)\right)\delta\psi + \kappa^2\mathrm{sech}^2(\kappa x)\delta\psi^*\right]
\end{align}
which with the expansion (\ref{BdG}) of $\delta\psi$ in $u(x),v(x)$ normal modes implies the Bogoliubov-de Gennes equations
\begin{align}\label{BdGV1}
    \Omega\left(\begin{matrix} u \\ -v\end{matrix}\right) &= -\frac{\hbar}{2M}\left(\begin{matrix} \frac{d^2}{dx^2} +\kappa^2 - \frac{x^2}{a_0^4} +4\kappa^2\mathrm{sech}^2(\kappa x) & 2\kappa^2\mathrm{sech}^2(\kappa x) \\ 2\kappa^2\mathrm{sech}^2(\kappa x) & \frac{d^2}{dx^2} +\kappa^2 - \frac{x^2}{a_0^4} +4\kappa^2\mathrm{sech}^2(\kappa x)\end{matrix}\right)\left(\begin{matrix} u \\ v\end{matrix}\right)\;.
\end{align}
\end{widetext}
We solve these equations separately in the two regions $|x|<L$ (the \emph{Soliton zone}) and $|x|>L$ (the \emph{Trap zone}); we find $u(x)$ and $v(x)$ in each zone, and match the two solutions smoothly at $x=\pm L$. By matching the solutions smoothly at $x=\pm L$, order by order in $\sqrt{\varepsilon}$ regardless of whether the expansion in $\varepsilon$ is an expansion in $\sqrt{2M\omega/\hbar}x$ or in $1/(\kappa x)$, we will ensure that the precise value of $L$ never appears in any of our final conclusions, because the $|x|<L$ solutions simply coincide with the $|x|>L$ solutions over the entire intermediate regions around $x=\pm L$.

\subsection{C.1: Index notation}
With the trapping confinement we will find a discrete set of normal modes, $(u_k,v_k)\to (u_\nu,v_\nu)$ and $\Omega(k)\to\Omega_\nu$ for discrete index $\nu$ instead of the continuous $k$ of the untrapped problem. The quantization of $\nu$ is due physically to the trapping potential, but we will determine it mathematically in the matching around $|x|\sim L$. 

In addition to the discrete index $\nu$ we will use a parity index, $(u_{\nu\pm},v_{\nu\pm})$. This is because, with the background bright soliton $\Psi_0$ in the center of the trap, and with a spatially even initial perturbation $\delta\psi(x,0)$ of (\ref{deltabr}), our Bogoliubov-de Gennes problem has reflection symmetry, and the $(u,v)$ normal modes will all be parity eigenstates, either even or odd as functions of $x$. 
Taking this parity into account will simplify the problem of finding the discrete $\nu$, allowing us to find the $\nu$ values that work for odd and even modes separately. 

When we then order all the modes sequentially by increasing eigenfrequency $\Omega_\nu$, labelling the $\nu$ indices as $\nu_n$ for whole-number $n \geq 0$, we will find that the even-$n$ modes are all even-parity $+$ modes, while the odd-$n$ modes are all odd. In this sense the $\pm$ parity labels will end up being superfluous, once we have found the discrete normal modes for all $n$ in sequence, and so we will ultimately stop using the $\nu$ and $\pm$ labels, and switch to using just $n$, which encodes $\pm = (-1)^n$ as well as the $\nu_n$ that we will determine below.

We hope that using $(u,v)$ with three different sets of indices $k$, $\nu\pm$, and $n$ in our paper is not too confusing! The continuous $k$ indices are known for the untrapped bright soliton, and the final $n\in \{0,1,2,\dots\}$ are inevitable for the confined system; the intermediate $\nu\pm$ notation provides the simplest path between $k$ and $n$ by matching the known Soliton zone solutions (\ref{Kaup}) with solutions in the Trap zone.

\subsection{C.2: Trap zone}
For $|x|>L$ we can neglect the $\mathrm{sech}^2(\kappa x)$ terms in (\ref{BdGV1}). This immediately decouples $u$ and $v$, and makes each of them separately solve a time-independent Schr\"odinger equation for a harmonic oscillator. We will find that a complete set of Bogoliubov-de Gennes normal modes is formed from $v$ that are all negligible in the Trap zone, so we will only need the solutions for $u$. They will exist for a discrete set of $\Omega\to\Omega_\nu$, and because they are solutions to a harmonic oscillator Schr\"odinger equation, we will use a mode label $\nu$ that is analogous to the harmonic oscillator energy quantum number $n$, with
\begin{equation}\label{Omeganu}
\Omega_\nu = \frac{\hbar\kappa^2}{2M}+\omega(\nu+1/2)\;.
\end{equation}
Because we are not only solving a harmonic oscillator Schr\"odinger problem, but rather the Bogoliubov-de Gennes equations with a bright soliton in the Soliton zone, we will find that $\nu$ is discrete but does not take exactly whole-number values. Determining the set of $\nu$ values, at least for $\varepsilon\ll 1$, is our main result in this paper.

The time-independent Schr\"odinger equation with a harmonic potential has parabolic cylinder functions as its eigenfunctions. For general real $\nu$ these functions all either decay as Gaussians for $x\to+\infty$ and blow up as inverse Gaussians for $x\to-\infty$, or else blow up and decay in the opposite directions of $x$. In the special case of whole-number $\nu$, there is one parabolic cylinder function which blows up in both directions, while the other decays in both directions and provides the square-integrable harmonic oscillator eigenfunction, equal to a Gaussian times a Hermite polynomial. Since we can use separate solutions in the separate regions $x>L$ and $x<-L$, however, we can allow non-integer $\nu$ while maintaining normalizability, just by taking parabolic cylinder functions that decay at large positive $x$ in the zone $x>L$, and functions that decay at large negative $x$ in the zone $x<-L$. 

In particular we can write for our even and odd solutions in the Trap zone
\begin{equation}
 |x|>L:\qquad\begin{matrix}   u_{\nu+} = Z^{T}_{\nu+}f_\nu\Big(\sqrt{\frac{2M\omega}{\hbar}}|x|\Big)\\
 u_{\nu-} =Z^{T}_{\nu-}\mathrm{sgn}(x)f_\nu\Big(\sqrt{\frac{2M\omega}{\hbar}}|x|\Big)\end{matrix}
\end{equation}
for parabolic cylinder functions $f_\nu$ and Trap-zone normalization factors $Z^{T}_{\nu\pm}$. To see how these solutions behave at $x=\pm L$, we can then use the small-argument limit of the parabolic cylinder functions to recognize
\begin{align}\label{ABT}
   u_{\nu\pm}(x) &\underset{x\to L^+}{\longrightarrow} Z^{T}_{\nu\pm}\frac{\sqrt{2\Gamma(\nu+1)}}{\big((2\nu+1)\pi\big)^{\frac{1}{4}}}\nonumber\\
   &\times \left(X(\nu)\cos\Big(\frac{\pi\nu}{2}\Big) \cos\Big(\sqrt{\nu+\frac{1}{2}}\sqrt{\frac{2M\omega}{\hbar}}x\Big)\right.\nonumber\\
&\quad \left.+\frac{1}{X(\nu)}\sin\Big(\frac{\pi\nu}{2}\Big) \sin\Big(\sqrt{\nu+\frac{1}{2}}\sqrt{\frac{2M\omega}{\hbar}}x\Big)\right)\nonumber\\
    X(\nu) &:= \sqrt{
    \frac{\Gamma(\frac{1+\nu}{2})}
    {\Gamma(1+\frac{\nu}{2})}
    }
    \Big(\frac{\nu+\frac{1}{2}}{2}\Big)^{\frac{1}{4}}\underset{\nu\gtrsim 2}{\longrightarrow}1
    \;.
\end{align}
Numerical plotting will confirm that these small-argument approximations are extremely accurate for $\sqrt{2M\omega/\hbar}x<1$, even for $\nu\sim 1$, and only become more accurate for larger $\nu$. The approximations smoothly interpolate between a simple Taylor series for small $\nu$ and the WKB approximation at large $\nu$. 

\begin{figure}
    \centering
    \includegraphics[width=0.8\linewidth]{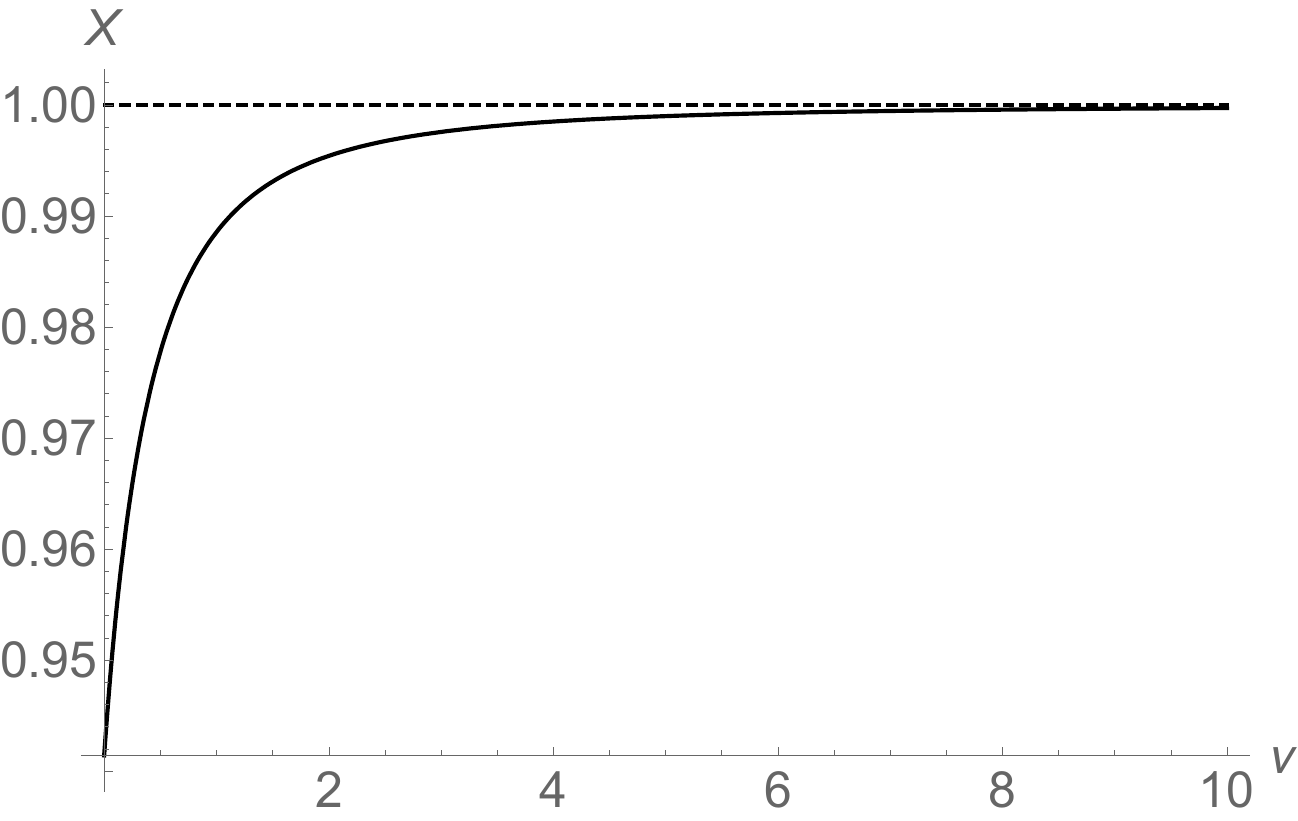}
    \caption{The factor $X(\nu)$ that appears in the small-argument limit of the parabolic cylinder functions in Eqn.~(\ref{ABT}) approaches the limit 1 quickly in $\nu$.}
    \label{Xfig}
\end{figure}

It is also convenient to note the numerical fact, shown in Fig.~\ref{Xfig}, that the factor $X(\nu)$ approaches 1 very quickly in $\nu$, so that the approximation $X(\nu)\to 1$ becomes excellent even just for $\nu\gtrsim 2$. Since our breathing modes will turn out to be composed of many eigenmodes, mostly with $\nu\gg 1$, we will replace $X(\nu)\to 1$ from now on.

The overall Trap-zone normalization factors $Z^T_{\nu\pm}$ must still be determined, by normalizing the entire $u_{\nu\pm}(x)$ that are defined for all $x$.

\subsection{C.3: Soliton zone}
Inside the Soliton zone $|x|<L$ we can apply our $V\to0$ solutions (\ref{Kaup}) for $u_k$ and $v_k$, with $k\to k(\nu)$ determined by
\begin{align}\label{knu}
    \Omega\big(k(\nu)\big) &= \frac{\hbar}{2M}\left(\kappa^2+k^2(\nu)\right) = \Omega_\nu \nonumber\\
    \Longrightarrow \quad k(\nu) &= \sqrt{\frac{2M\omega}{\hbar}(\nu+\frac{1}{2})}= \kappa \varepsilon\sqrt{\nu+\frac{1}{2}}\;.
\end{align}
We can form even- and odd-parity solutions as
\begin{equation}
    \left(\begin{matrix}u_{\nu\pm}\\ v_{\nu\pm}\end{matrix}\right) = Z_{\nu\pm}^{S}\left(\begin{matrix}u_{k(\nu)}\pm u_{-k(\nu)}\\ v_{k(\nu)}\pm v_{-k(\nu)}\end{matrix}\right)
\end{equation}
for normalization factors $Z_{\nu\pm}^{S}$.

We can also recognize that the time-translation zero mode $R_{+},S_+$ remains in the presence of the harmonic trap, with eigenfrequency zero, while the spatial translation zero mode $R_-,S_-$ of the trap-free case becomes the Kohn dipole mode in the trap, with $\omega_d = \omega$ exactly. Neither of these modes extends non-negligibly into the Trap zone, and both modes remain well separated in frequency from the modes with $\Omega_\nu\geq \frac{\hbar\kappa^2}{2M}$.

To compare these Soliton zone solutions with the Trap zone solutions in the border region $|x|\sim L$, we can first note that the Soliton zone $v_{\nu\pm}$ do all vanish rapidly for $|x|\gtrsim L$, and thus trivially match the vanishing Trap zone $v_{\nu\pm}$. As we approach $|x|=L$ from inside the Soliton zone, the Soliton-zone solutions for $u_{\nu\pm}$ do not vanish but instead approach
\begin{align}\label{ABS}
    &u_{\nu\pm}(x) \underset{x\to L^-}{\longrightarrow} Z^{S}_{\nu\pm}\left(A_{\nu}^\pm \cos\Big(\sqrt{\nu+\frac{1}{2}}\sqrt{\frac{2M\omega}{\hbar}}x\Big)\right.\nonumber\\
&\qquad\qquad\qquad\left.+B_{\nu}^\pm \sin\Big(\sqrt{\nu+\frac{1}{2}}\sqrt{\frac{2M\omega}{\hbar}}x\Big)\right)\nonumber\\
    &A^{+}_{\nu} = -iB^{-}_{\nu}  = \sqrt{\frac{2}{\pi}}\frac{k^2(\nu)-\kappa^2}{k^2(\nu)+\kappa^2}=\sqrt{\frac{2}{\pi}}\frac{\varepsilon^2(\nu+\frac{1}{2})-1}{\varepsilon^2(\nu+\frac{1}{2})+1} \nonumber\\
    &B^{+}_{\nu} = iA^{-}_{\nu} =-\sqrt{\frac{2}{\pi}}\frac{2\kappa k(\nu)}{k^2(\nu)+\kappa^2}=-\sqrt{\frac{2}{\pi}}\frac{2\varepsilon\sqrt{\nu+\frac{1}{2}}}{\varepsilon^2(\nu+\frac{1}{2})+1}\;.
\end{align}

Because for small $\varepsilon$ the two different approximations $\mathrm{sech}(\kappa x)\ll 1$ and $\varepsilon\kappa x\ll 1$ are both simultaneously valid near $|x|\sim L$, the Soliton-zone and Trap-zone solutions are both accurate approximations for $u(x)$ and $v(x)$ in this border region, and so it is no accident that they have the $u_{\nu\pm}$ in the two zones have the same sinusoidal forms for $|x|\sim L$. To make the solutions match smoothly, therefore, we only need to tune their respective coefficients together.

\subsection{C.4: Matching and quantization of $\nu$}
We combine our separate Soliton-zone and Trap-zone solutions into global solutions for all $x$ by requiring (\ref{ABT}) and (\ref{ABS}) to agree for $x\sim \pm L$. This means setting
\begin{equation}\label{match1}
Z^{S}_{\nu\pm}\left(\begin{matrix}A^{\pm}_{\nu}\\ B^{\pm}_{\nu}\end{matrix}\right) = Z^{T}_{\nu\pm}\frac{\sqrt{2\Gamma(\nu+1)}}{\big((2\nu+1)\pi\big)^{\frac{1}{4}}}\left(\begin{matrix}\cos\Big(\frac{\pi\nu}{2}\Big)\\ \sin\Big(\frac{\pi\nu}{2}\Big)\end{matrix}\right)\;,
\end{equation}
when we make the excellent approximation $X(\nu)\to 1$ as discussed above.

For our main result in the next section, it will be necessary to determine the normalization factors $Z_\nu^{S,T}$ themselves. This we do in Appendix B; we will quote and use the convenient result in the next section of the paper. To obtain the quantization condition on $\nu$, however, we can simply eliminate the $Z_\nu$ factors from (\ref{match1}), by dividing the lower line of (\ref{match1}) by the upper line. This yields the quantization condition 
\begin{equation}\label{quant0}
    \tan\Big(\frac{\pi\nu}{2}\Big) = \frac{B_\nu^{\pm}}{A_\nu^{\pm}}\;,
\end{equation}
which are two different transcendental equations (for the even and odd cases $\pm$) with solutions for infinite discrete sets of $\nu$. Since the tangent on the left side of (\ref{quant0}) sweeps through all real numbers within every interval $n<\nu<n+2$ for integer $n$, both the even and odd sets of eigenfrequencies will be discrete but infinite, and will each have one eigenfrequency in almost every interval $n<\nu<n+2$---the exception being the interval in which $\varepsilon^2(\nu+1/2)$ passes through one. 

We can simplify (\ref{quant0}) to make the discrete spectrum of $\nu$ clearer. Firstly we can exploit the fact that $B_\nu^-/A_\nu^- = - A_\nu^+/B_\nu^+$, together with some trigonometry, by writing
\begin{equation}\label{nu}
\nu = n + \Delta_n
\end{equation}
for whole $n$, where even $n$ yield the $\nu$ for even modes $u_{\nu+}$ and odd $n$ give the odd modes $u_{\nu-}$. This conveniently implies the same equation for $\Delta_n$ for both even and odd modes,
\begin{equation}\label{deltan1}
    \Delta_n = \frac{2}{\pi}\tan^{-1}\left(\frac{2\varepsilon\sqrt{n+\Delta_n+1/2}}{1-\varepsilon^2(n+\Delta_n+1/2)}\right)\;.
\end{equation}
We can then apply the identity
\begin{equation}
    \cos^2(\theta) = \frac{1}{1+\tan^2(\theta)}\;,
\end{equation}
and eliminate spurious roots by carefully considering the sign of $\tan(\pi\Delta_n/2)$, to reach the smoother-looking but equivalent equation
\begin{equation}\label{deltan2}
    \Delta_n = \frac{2}{\pi}\cos^{-1}\left(\frac{1-\varepsilon^2(n+\Delta_n+1/2)}{1+\varepsilon^2(n+\Delta_n+1/2)}\right)\;.
\end{equation}
Since all the $\Delta_n$ are obviously of order unity, for small $\varepsilon^2$ we can always neglect $\Delta_n+1/2$ on the right side of (\ref{deltan1}), even when $n$ is so large that $\epsilon^2 n$ may not be small. This final very accurate approximation provides the compact, closed result that we use as (\ref{deltan3}) in our main text.

\section{Appendix D: Normalization}

While our quantization condition (\ref{quant0}) for $\nu\to\nu_n$ was revealed by dividing the two lines of Eqn.~(\ref{match1}), we can find the relation between the Soliton-zone and Trap-zone normalization factors $Z_{\nu\pm}^{S}$ and $Z_{\nu\pm}^{T}$ by modulus-squaring the two lines of (\ref{match1}) and adding them. This yields
\begin{align}\label{STZ}
    \Big|Z_{\nu\pm}^{T}\Big|^2 &= \frac{\sqrt{(2\nu+1)}}{\sqrt{\pi}\Gamma(\nu+1)}\Big|Z_{\nu\pm}^{S}\Big|^2\;.
\end{align}

For the discrete analog of the delta-function normalization in the untrapped case, we now need to impose the finite normalization condition
\begin{equation}\label{normcond}
    \int_{-\infty}^\infty\!dx\, \Big(|u_{\nu\pm}|^2-|v_{\nu\pm}|^2\Big) \equiv 2\int_{0}^\infty\!dx\, \Big(|u_{\nu\pm}|^2-|v_{\nu\pm}|^2\Big) = 1\;,
\end{equation}
where we use the fact that $u_{\nu\pm}$ and $v_{\nu\pm}$ are either even or odd, so that their product is an even function of $x$.
This condition (\ref{normcond}) applies only for the discrete cases $\nu\to\nu_n$ and $\pm\to(-1)^n$ that are actually in our spectrum of modes; it will finally fix $Z^{S}_{\nu_n,(-1)^n}$, which we will re-write as $Z_n$ for each discrete mode. We compute the integral separately in the Trap and Soliton zones and add the results together; using the fact that $v_{\nu\pm}\to0$ in the Trap zone, and also applying (\ref{STZ}), this provides
\begin{widetext}\begin{align}\label{Sigmadef}
    \frac{1}{2} &=\int_0^L\!dx\, \Big(|u_{\nu\pm}|^2-|v_{\nu\pm}|^2\Big) + \int_L^\infty\!dx\, |u_{\nu\pm}|^2=:\big|Z_{n}\big|^2 \Big(\Sigma^S_{n} +\Sigma^T_{n} \Big)\nonumber\\
\Sigma^S_{n} &= \int_0^L\!dx\,\Big(|u_{k(\nu_n)}+(-1)^n u_{-k(\nu_n)}|^2-|v_{k(\nu_n)}+(-1)^n v_{-k(\nu_n)}|^2\Big)\nonumber\\
\Sigma^T_{n} & = \frac{\sqrt{(2\nu_n+1)}}{\sqrt{\pi}\Gamma(\nu_n+1)}\int_L^\infty\! dx\,\Big[f_{\nu_n}\Big(\sqrt{\frac{2M\omega}{\hbar}}x\Big)\Big]^2= \frac{\sqrt{(2\nu_n+1)}}{\sqrt{\pi}\Gamma(\nu_n+1)}\sqrt{\frac{\hbar}{2M\omega}}\int_{\sqrt{\frac{2M\omega}{\hbar}}L}^\infty\! dz\,[f_{\nu_n}(z)]^2
\;.
\end{align}

To compute $\Sigma^S_n$ we group terms in the explicit forms for $u_{\pm k(\nu)}$ and $v_{\pm k(\nu)}$ as given by (\ref{Kaup}) with $k(\nu)=\varepsilon\kappa\sqrt{\nu+1/2}$, to reveal
\begin{align}
  \Sigma_{n}^S &= \frac{1}{\pi\big(1+\varepsilon^2(\nu+1/2)\big)^2}\int_0^L\!dx\,\Bigg\{\Big(\varepsilon^2(\nu+1/2)+1\Big)^2+(-1)^n
 \Big(\varepsilon^4(\nu+1/2)^2-6\varepsilon^2(\nu+1/2)+1\Big)\cos(2kx)\nonumber\\
     &\qquad -\frac{2}{\kappa}\frac{d}{dx}\Big[\tanh(\kappa x)\big((1+\varepsilon^2(\nu+1/2))+(-1)^n\big(1-\varepsilon^2(\nu+1/2)\big)\cos(2kx)\big)-(-1)^n \varepsilon\sqrt{\nu+1/2}\,\mathrm{sech}^2(\kappa x)\sin(2kx) 
\Big]
  \Bigg\}\;,
\end{align}
where we omit the $_n$ subscripts on $\nu$ for compactness.
This is straightforward to integrate:
\begin{align}\label{sigmaS}
    \Sigma^S_{n} = \frac{1}{\pi}\left(L - \frac{2}{\kappa[1+\varepsilon^2(\nu+\frac{1}{2})]} +(-1)^n \frac{1-6\varepsilon^2(\nu+\frac{1}{2})+\varepsilon^4(\nu+\frac{1}{2})^2}{2\kappa\varepsilon\sqrt{\nu+\frac{1}{2}}[1+\varepsilon^2(\nu+\frac{1}{2})]^2}\sin(2kL)-2(-1)^n\frac{1-\varepsilon^2(\nu+\frac{1}{2})}{[1+\varepsilon^2(\nu+\frac{1}{2})]^2}\cos(2kL)\right)\;.
\end{align}
This long expression for general real $\nu$ simplifies considerably in the special cases of our discrete eigenvalues $\nu\to\nu_n$ as given by (\ref{nu}) and (\ref{deltan1}). Using (\ref{nu}) and (\ref{deltan1}) we find
\begin{align}
    \cos(\pi\nu_n) &= (-1)^n \frac{1-6\varepsilon^2(\nu_n+\frac{1}{2})+\varepsilon^4(\nu_n\frac{1}{2})^2}{2\kappa\varepsilon\sqrt{\nu_n+\frac{1}{2}}[1+\varepsilon^2(\nu_n+\frac{1}{2})^2]^2}\nonumber\\
    \sin(\pi\nu_n) &= (-1)^n \frac{4\varepsilon\sqrt{\nu_n+\frac{1}{2}}[1-\varepsilon^2(\nu_n+\frac{1}{2}]}{[1+\varepsilon^2(\nu_n+\frac{1}{2})^2]^2}\;,
\end{align}
so that for $\nu=\nu_n$ we can re-write (\ref{sigmaS}) as
\begin{align}\label{sigmaS2}
    \Sigma^S_{n} = \frac{1}{\pi}\left(L - \frac{2}{\kappa[1+\varepsilon^2(\nu_n+\frac{1}{2})]}+\frac{\sin\big(2k(\nu_n)L-\pi\nu_n\big)}{2\kappa\varepsilon\sqrt{\nu_n+\frac{1}{2}}}\right)\;.    
\end{align}

Then on the other hand we can evaluate $\Sigma^T_{n}$ by combining the parabolic cylinder equation itself,
\begin{equation}\label{parsol}
   \nu f_\nu(z) = -f_\nu''(z) + \left(\frac{z^2}{4}-\frac{1}{2}\right)f_\nu(z)\;,
\end{equation}
with its derivative with respect to $\nu$,
\begin{align}\label{parsolnu}
    f_\nu = \Big(-\frac{\partial^2}{\partial z^2}+\frac{z^2}{4}-\frac{1}{2}-\nu\Big)\frac{\partial f_\nu}{\partial \nu}\;.
\end{align}
First we use the parabolic cylinder equation (\ref{parsol}) to write trivially
\begin{align}\label{partriv}
     [f_\nu(z)]^2  &= [f_\nu]^2-\left[-f_\nu'' + \left(\frac{z^2}{4}-\frac{1}{2}-\nu\right)f_\nu\right]\frac{\partial f_\nu}{\partial\nu}\;,
\end{align}
since the whole term in large square brackets is just zero, by (\ref{parsol}). Then we multiply (\ref{parsolnu}) by $f_\nu(z)$:
\begin{align}\label{parmult}
    [f_\nu]^2 = f_\nu\Big(-\frac{\partial^2}{\partial z^2}+\frac{z^2}{4}-\frac{1}{2}-\nu\Big)\frac{\partial f_\nu}{\partial \nu}\;.
\end{align}
Inserting (\ref{parmult}) in (\ref{partriv}) lets all the terms without $z$-derivatives cancel, leaving
\begin{align}
    [f_\nu(z)]^2  &=\frac{\partial^2f_\nu}{\partial z^2}\frac{\partial f_\nu}{\partial \nu}-f_\nu\frac{\partial^3f_\nu}{\partial\nu\partial z^2} \equiv \frac{\partial}{\partial z}\left(\frac{\partial f_\nu}{\partial z}\frac{\partial f_\nu}{\partial \nu}-f_\nu\frac{\partial^2f_\nu}{\partial\nu\partial z}\right)\;.
\end{align}
This exact expression of $f_\nu^2$ as a derivative with respect to $z$ makes the integral $\Sigma^T_n$ straightforward in spite of the complexity of the parabolic cylinder functions of general order $\nu$.
Since the $f_\nu(z)$ also all vanish rapidly for $z\to\infty$, we therefore obtain
\begin{align}
    \Sigma^T_{n} &=  -\frac{\sqrt{(2\nu+1)}}{\sqrt{\pi}\Gamma(\nu+1)}\sqrt{\frac{\hbar}{2M\omega}}\left(\frac{\partial f_\nu}{\partial z}\frac{\partial f_\nu}{\partial \nu}-f_\nu\frac{\partial^2f_\nu}{\partial\nu\partial z}\right)\Big|_{z\to\sqrt{\frac{2M\omega}{\hbar}}L}\;.
\end{align}
Then using the small-argument limit of $f_\nu(z)$ as in (\ref{ABT}) with the excellent approximation $X(\nu)\to1$ as explained in Fig.~3, 
\begin{equation}
    f_\nu(z) \doteq \frac{\sqrt{2\Gamma(\nu+1)}}{\Big((2\nu+1)\pi\Big)^{\frac{1}{4}}}\cos\Big(\frac{\pi\nu}{2}-\sqrt{\nu+\frac{1}{2}}z\Big)\;,
\end{equation}
we obtain
\begin{align}\label{sigmaT}
    \Sigma^T_{n} &= \sqrt{\frac{\hbar}{2M\omega}}\sqrt{\nu_n+\frac{1}{2}}-\frac{1}{\pi}\left(L+\frac{\sin\big(2k(\nu_n)L-\pi\nu_n\big)}{2\kappa\varepsilon\sqrt{\nu+\frac{1}{2}}}\right)\;,
\end{align}
using $\kappa\varepsilon = \sqrt{2M\omega/\hbar}$.

Combining (\ref{sigmaS2}) and (\ref{sigmaT}), we see that in the sum $\Sigma_n^S+\Sigma_n^T$ all terms dependent on $L$ cancel each other, so that from (\ref{Sigmadef}) we have
\begin{align}\label{Znfin}
    |Z_n|^2 &= \frac{1}{2(\Sigma_n^S+\Sigma_n^T)} = \frac{1}{2}\sqrt{\frac{2M\omega}{\hbar}}\frac{1}{\sqrt{\nu_n+\frac{1}{2}}-\frac{2\varepsilon}{\pi}\frac{1}{1+\varepsilon^2(\nu_n+\frac{1}{2})} }\nonumber\\
    &= \frac{1}{2\sqrt{\nu_n+\frac{1}{2}}}\sqrt{\frac{2M\omega}{\hbar}}\left(1+\frac{2}{\pi}\frac{\varepsilon}{\sqrt{\nu_n+\frac{1}{2}}[1+\varepsilon^2(\nu_n+\frac{1}{2})]}+\mathcal{O}(\varepsilon^2)\right)\equiv \frac{dk(\nu_n)}{dn}\;.
\end{align}
In the final step, here, we use $k(\nu) = \sqrt{2M\omega/\hbar}\sqrt{\nu+1/2}$ as always, and use $\nu=n+\Delta_n$ as given by (\ref{deltan1}), in order to differentiate $k(\nu_n)$ with respect to $n$ as if $n$ were a continuous variable.

The final line of Eqn.~(\ref{Znfin}) is the convenient result which allows us to approximate sums over $n$ with integrals in our main text. In hindsight this result may appear inevitable; it could perhaps have been guessed without going through this complicated appendix. We know of no simpler way than our analysis here, however, to confirm that this really is true.

As a small reward for any readers who have followed to this point, we can point out that combining (\ref{Znfin}) with (\ref{STZ}) lets us confirm that for $n\to\infty$ the high-frequency wave function $u_n(x)$ from (\ref{BdG}) approaches
\begin{equation}
   \lim_{n\to\infty} u_n(x) = \left(\frac{M\omega}{\hbar\pi}\right)^{\frac{1}{4}}\frac{f_n\big(\sqrt{\frac{2M\omega}{\hbar}}x\big)}{\sqrt{n!}}
\end{equation}
in the Trap zone, which is the familiar harmonic oscillator result that we expect at high frequencies, when the soliton background becomes a negligible perturbation compared to the excitation's huge single-particle energy.
\end{widetext}

\end{document}